\documentclass[11pt,a4paper]{article}
\pdfoutput=1
\usepackage{jheppub}
\usepackage[utf8]{inputenc}

\usepackage{color}
\usepackage{amsmath}
\usepackage{pifont}
\usepackage{bbold}

\usepackage{bbm}
\usepackage{verbatim}   
\usepackage{subfigure}
\usepackage{acronym}

\usepackage{amsfonts}
\usepackage{amssymb}
\usepackage{mathrsfs}
\usepackage{graphicx}
\usepackage{multirow}
 \usepackage{slashed}

\usepackage{soul}

\usepackage{caption}
\usepackage{tikz}
\usetikzlibrary{decorations.pathreplacing}

\graphicspath{{./fig/}}

\definecolor{mypurple}{RGB}{164,64,214}

\newcommand{\zz}{\mathbb{Z}}
\newcommand{\nn}{\mathbb{N}}
\newcommand{\rr}{\mathbb{R}}

\newcommand{\der}{\mathrm{d}}

\def\bar#1{\overline{#1}}

\def\abs#1{\left| #1\right|}

\def\inv{^{\raise.15ex\hbox{${\scriptscriptstyle -}$}\kern-.05em 1}}
\def\lbar{{\lower.35ex\hbox{$\mathchar'26$}\mkern-10mu\lambda}} 

\def\to{\rightarrow}

\let\<=\langle
\let\>=\rangle

\let\+=\uparrow

\newcommand{\newc}{\newcommand}
\newc{\gsim}{\lower.7ex\hbox{$\;\stackrel{\textstyle>}{\sim}\;$}}
\newc{\lsim}{\lower.7ex\hbox{$\;\stackrel{\textstyle<}{\sim}\;$}}

\title{Disassembling the Clockwork Mechanism}

\author[a]{Nathaniel Craig,}
\emailAdd{ncraig@physics.ucsb.edu}
\author[b,c]{Isabel Garcia Garcia}
\emailAdd{isabel.garciagarcia@physics.ox.ac.uk}
\author[a]{and Dave Sutherland}
\emailAdd{dwsuth@ucsb.edu}

\affiliation[a]{Department of Physics, University of California, Santa Barbara, CA 93106, USA}
\affiliation[b]{Rudolf Peierls Centre for Theoretical Physics, University of Oxford, Oxford, OX1 3NP, UK}
\affiliation[c]{Kavli Institute for Theoretical Physics, University of California, Santa Barbara, CA 93106, USA}

\abstract{The clockwork mechanism is a means of naturally generating exponential hierarchies in theories without significant hierarchies among fundamental parameters. We emphasize the role of interactions in the clockwork mechanism, demonstrating that clockwork is an intrinsically abelian phenomenon precluded in non-abelian theories such as Yang-Mills, non-linear sigma models, and gravity. We also show that clockwork is not realized in extra-dimensional theories through purely geometric effects, but may be generated by appropriate localization of zero modes.
}

\begin{document}

\maketitle


\section{Introduction}
\label{sec:intro}

The problems of the Standard Model remain as striking as ever, but their solutions --- if they indeed exist --- have yet to make themselves apparent. From the electroweak hierarchy problem to the dark matter puzzle to the inflationary paradigm, experimental data largely disfavors solutions involving mass scales and couplings commensurate with those seen elsewhere in nature. 

Perhaps this is a sign that the degrees of freedom solving the problems of the Standard Model are in some way sequestered from us, interacting feebly due to small dimensionless couplings or the suppression by vast dimensionful scales. Indeed, extensions of the Standard Model operating along these lines are among the most compatible with existing data: cosmological observations are accommodated by inflationary potentials that are flat on trans-Planckian scales; the electroweak hierarchy problem may be solved by the evolution of fields across similarly trans-Planckian distances \cite{Graham:2015cka}; and dark matter may be explained by light particles carrying infinitesimal electromagnetic charges. Recent attempts to test these feebly-interacting degrees of freedom have led to a proliferation of novel experiments across the energy, intensity, and cosmic frontiers.

Such feeble interactions require large parametric hierarchies with respect to the couplings and scales of the Standard Model and quantum gravity. These parametric hierarchies are challenging to understand from the perspective of naturalness, which prefers $\mathcal{O}(1)$ dimensionless couplings and degenerate scales in the fundamental theory. Even parameters that are technically natural or otherwise radiatively stable  beg for deeper explanation if they are infinitesimally small. Beyond questions of field-theoretic naturalness, extremely weak couplings are challenging to reconcile with generic properties of quantum gravity \cite{ArkaniHamed:2006dz}.

 To this end, there has recently been considerable progress in generating large effective hierarchies from theories whose fundamental parameters are all natural in the conventional sense. These include models of inflation with sub-Planckian intrinsic scales and super-Planckian effective couplings \cite{Kim:2004rp,Choi:2014rja,delaFuente:2014aca}, as well as more general theories realizing exponential hierarchies in the decay constants of pseudo-goldstone bosons~\cite{Choi:2015fiu,Kaplan:2015fuy}. Such ``clockwork'' models involve a linear quiver with $N+1$ sites, where each site possesses a global $U(1)$ symmetry acting on a complex scalar field. The $U(1)^{N+1}$ symmetry of the quiver is explicitly broken by asymmetric nearest-neighbor interactions that preserve a single $U(1)$. When the scalars acquire vacuum expectation values, the resulting goldstone boson is a linear combination of fields from each site whose weights follow a geometric sequence, and the unbroken symmetry is asymmetrically distributed among sites. As a result, any coupling of additional fields to the scalar at a specific site gives rise to an exponentially-suppressed and site-dependent coupling of those fields to the goldstone boson. This provides a natural mechanism for generating exponential hierarchies in a theory whose fundamental parameters are all of comparable size,
and leads to a variety of model-building possibilities \cite{Farina:2016tgd,Hambye:2016qkf,Coy:2017yex}.
    
In~\cite{CW}, the clockwork mechanism was generalized to include states of higher spin, giving rise to exponentially small fermion masses, gauge millicharges, and gravitational couplings. Even more ambitiously, the authors of~\cite{CW} also conjecture a continuum counterpart to four-dimensional clockwork in the form of five-dimensional linear dilaton models, which in turn are holographically related (with the addition of two more compact dimensions) to little string theory \cite{Aharony:1998ub}. If true, this would open the door to a wider variety of constructions in both four and five dimensions \cite{Kehagias:2016kzt,Ahmed:2016viu}.

Given the potentially vast applications of clockwork to questions of phenomenological interest, it is crucial to precisely determine the scope of clockwork. As such, in this paper we systematically answer two questions:
{\it
\begin{enumerate}
\item What theories can be clockworked in four dimensions?
\item What are their higher-dimensional continuum counterparts?
\end{enumerate} 
}
To answer these questions, we must take care to carefully define the features of a clockwork theory.
In particular, the definition must distinguish genuine ``clockwork'' phenomena from already-familiar hierarchies arising from volume suppression \cite{ArkaniHamed:1998rs} or curvature-induced localization \cite{Randall:1999ee} in extra dimensions (or their deconstructed counterparts \cite{ArkaniHamed:2001ca, Hill:2000mu, Randall:2002qr}). For our purposes, we will take clockwork to involve the salient features of the original models \cite{Choi:2015fiu,Kaplan:2015fuy}, namely

\begin{quotation}
\noindent {\bf Clockwork:} {\it A four-dimensional quiver theory with no exponential hierarchies in fundamental parameters that gives rise to exponentially suppressed (and site-dependent) couplings to a symmetry-protected zero mode.}
\end{quotation}

These are not merely incidental properties of clockwork, but essential ones. In particular, site-dependent exponentially suppressed couplings are a hallmark of the asymmetric distribution of the unbroken symmetry among different sites. This clearly distinguishes the clockwork theories of \cite{Choi:2015fiu,Kaplan:2015fuy} in four dimensions from, say, deconstructions of extra dimensions with flat or bulk AdS metrics.
For example, deconstructions of flat extra dimensions involve no hierarchies in fundamental parameters, but only give rise to site-independent zero mode couplings suppressed by $\sim \sqrt{N}$ factors. Similarly, deconstructions of Randall-Sundrum and other warped models can give rise to exponentially-suppressed (albeit position-independent) zero mode couplings, but necessarily involve exponential hierarchies in the vacuum expectation values of the link fields. The genuine novelty of clockwork is that it furnishes exponential and site-dependent effective couplings from a fundamental theory with no large parametric hierarchies or multiplicity of sites. To the extent that these properties arise from the asymmetric distribution of an unbroken symmetry subgroup, in what follows we will refer to the localization of fields in the space of appropriate symmetry generators as `symmetry-localization.' Such symmetry-localization controls the couplings of fields dictated by gauge or global symmetries. As we will see, this symmetry-localization differs in important ways from localization of fields propagating in a non-trivial geometry with respect to a 5D metric.
 
As we will show, the answers to these questions are:
\begin{enumerate}
\item {\it Clockwork is a strictly abelian phenomenon.} In particular, there is no clockwork for Yang-Mills theories, non-linear sigma models, or gravity.
\item {\it Geometry alone cannot clockwork bosonic fields.} Zero modes of massless bosonic bulk fields are flat, regardless of apparent features of the metric. In particular, higher-dimensional models with massless bulk fields on linear dilaton backgrounds do not furnish continuum counterparts of clockwork. Successful continuum clockwork requires bulk and brane masses to symmetry-localize the zero mode.
\end{enumerate} 

These conclusions are consistent with the original clockwork proposals \cite{Choi:2015fiu,Kaplan:2015fuy}, but they are in tension with the results of \cite{CW},
applications thereof \cite{Kehagias:2016kzt}, and subsequent attempts to clockwork non-abelian global symmetries \cite{Ahmed:2016viu}.
Insofar as it is not possible to clockwork gravity in the sense of generating an asymmetrically-distributed general coordinate invariance, clockwork offers no new solution to the electroweak hierarchy problem.
Moreover, in those cases where clockwork is possible, namely for spin-0 and abelian spin-1 fields, we argue that --- appropriately interpreted --- deconstructions of five-dimensional linear dilaton models do not exhibit clockwork phenomena.

We emphasize that our statement about the lack of a clockwork solution to the hierarchy problem stems solely from the fact that gravity cannot be consistently
clockworked, as we prove in section~\ref{sec:discreteGraviton}.
This is \emph{not} a statement about the potential of linear dilaton theories for solving the hierarchy problem ---
that they do is well-known \cite{Antoniadis:2011qw,Antoniadis:2001sw}. In these theories exponential hierarchies are generated by a linear profile for the dilaton, whose exponential coupling gives rise to the desired hierarchies. When deconstructed, they do not lead to four-dimensional theories with a clockwork graviton in which the surviving general coordinate invariance is asymmetrically distributed among different sites.\footnote{Very much in the same
way that other extra dimensional solutions, like Randall-Sundrum \cite{Randall:1999ee} or large flat extra dimensions \cite{ArkaniHamed:1998rs},
do not lead to a clockwork graviton when deconstructed.}

More optimistically, we construct five-dimensional theories with bulk and brane masses that exhibit clockwork phenomena. These are the continuum counterparts of clockwork theories, in the sense that discretizing them gives four-dimensional theories whose spectra and couplings match those of a uniform four-dimensional clockwork up to appropriately small $1/N$ corrections. The emergence of meaningful clockwork phenomena in the deconstruction of higher-dimensional theories with bulk and brane masses opens the door to a variety of promising model-building possibilities.

We stress that further model building opportunities may arise if the definition of clockwork is significantly relaxed. In particular, if we do not require that the zero-mode be symmetry protected, it is possible to construct a quiver of non-linear sigma models whose zero-mode has exponentially suppressed, and site-dependent, couplings \cite{Fonseca:2016eoo}. Whilst such a zero-mode is necessarily massive, it may be parametrically lighter than the other modes of the quiver --- a fact which is mirrored in the quiver's 5D analog as the fifth component of a non-Abelian gauge field in AdS. However, in keeping with the original clockwork model, we will insist on a symmetry-protected, massless zero-mode in the rest of the paper.

The paper is organized as follows: In section \ref{sec:discrete} we review the essential features of the discrete clockwork mechanism, following the arguments of \cite{CW},
and illustrate how effective clockworking arises only for goldstone bosons of spontaneously broken abelian global symmetries, and gauge bosons of abelian gauge symmetries.
We explicitly show how an analogous mechanism cannot be built for non-abelian gauge bosons and gravitons.
We establish the abelian nature of clockwork more rigorously in section \ref{sec:GT} using group-theoretic arguments, independent of any specific quiver constructions.
In section \ref{sec:noMasses} we turn to the conjectured continuum counterpart of viable four-dimensional clockwork.
We show that the couplings between the zero mode of a massless bulk scalar or vector and matter localized at some point in the fifth dimension
do not reproduce the properties of clockwork models when deconstructed -- a statement that holds for a general class of warped metrics,
and includes linear dilaton theories.
Given the failure of geometry alone to produce clockwork, in section \ref{sec:flat} we show that genuine clockwork arises in the deconstruction of extra dimensions with a flat metric and suitably-chosen bulk and brane mass terms that preserve a massless zero mode. We summarize our conclusions in section \ref{sec:conclusions}, and reserve more general group-theoretic arguments in both four and five dimensions for appendix \ref{app:group}.
Finally, in appendix \ref{app:graviton} we explicitly show how the deconstruction of a gravitational extra dimension does not lead to a graviton clockwork,
in keeping with our results of section \ref{sec:discrete}.

\section{Discrete clockwork}
\label{sec:discrete}

In this section, we discuss the basic features of the discrete clockwork mechanism
using the framework introduced in~\cite{CW}.
Sections~\ref{sec:discreteScalar} and \ref{sec:discreteVector} focus on the spin-0, and abelian spin-1 scenarios,
in which a finite amount of clockworking may be successfully generated in a consistent fashion
(as defined in the Introduction).
On the other hand, sections~\ref{sec:discreteNonAbelian} and \ref{sec:discreteGraviton} illustrate how an analogous clockwork
mechanism cannot be consistently constructed in the non-abelian spin-1 and spin-2 cases. We summarize these results from the perspective of the clockwork symmetry in section \ref{sec:symmetries}.

\subsection{Scalar clockwork}
\label{sec:discreteScalar}

The discrete scalar clockwork mechanism involves $N+1$ real scalar fields, together with $N$ charge and mass-squared
parameters, $q_j$ and $m_j^2$ ($j=0,...,N-1$), such that the lagrangian of the scalar sector is given by\footnote{Throughout the document, implicit contraction of Greek indices denotes contraction with $\eta_{\mu \nu} = (-1,+1,+1,+1)$.}
\begin{equation}
	\mathcal{L}_4 = - \frac{1}{2} \sum_{j=0}^N (\partial_\mu \phi_j)^2 - \frac{1}{2} \sum_{j=0}^{N-1} m_j^2 (\phi_j - q_j \phi_{j+1})^2 \ .
\label{eq:L4scalarDiscrete}
\end{equation}

The $N+1$ scalar fields $\phi_j$ may be conveniently thought of as the Goldstone bosons of a global $U(1)^{N+1}$ symmetry, spontaneously
broken at some high scale $f$. Eq.(\ref{eq:L4scalarDiscrete}) can then be regarded as the effective lagrangian of the Goldstone sector,
valid at scales $\ll f$, and with the mass-squared parameters $m_j^2$ introducing an explicit breaking of $N$ of the $N+1$ global symmetries.
As a result, the effective theory of the Goldstone sector features only one massless state.

The parameters $m_j^2$ may arise from the vacuum expectation values (vev's)
of $N$ additional scalar fields charged under the $U(1)_j$ and $U(1)_{j+1}$ global subgroups, with charges $+1$ and $-q_j$ respectively,
as discussed in~\cite{CW}.
This allows for the effective theory defined through eq.(\ref{eq:L4scalarDiscrete}) to be UV completed in a way such that
all sources of symmetry breaking are spontaneous.

The profile of the massless mode corresponding to the single Goldstone that remains in the spectrum
is given by $\phi_{(0)} = \sum_{j=0}^N c_j \phi_j$, with
\begin{equation}	 
	c_j = c_0 \prod_{k=0}^{j-1} \frac{1}{q_k} \quad ({\rm for} \ j \geqslant 1) \ , \quad {\rm and} \quad
	c_0 = \left( 1 + \sum_{j=1}^{N} \prod_{k=0}^{j-1} \frac{1}{q_k^2} \right)^{-1/2} \ ,
\label{eq:discreteScalarProfile}
\end{equation}
where the expression for $c_0$ comes from demanding the kinetic term of $\phi_{(0)}$ be canonical.
In particular, in the case of equal clockwork parameters ($m_j^2 \equiv m^2, q_j \equiv q \ \forall \ j = 0,...,N-1$) considered in \cite{CW}, one finds $c_j \simeq q^{-j}$ (for $q>1$ and large $N$).
The massless mode therefore has a profile that is exponentially localized towards the $j=0$ site.

The clockwork mechanism as a means of generating large hierarchies comes into play when we introduce an axion-like coupling
between one of the scalar fields (e.g.~the scalar field of the $k$-th site), and a non-abelian gauge theory, of the form
\begin{equation}
	\mathcal{L}_4 \supset - \frac{1}{4 {\bar g}^2} G_{\mu \nu} G^{\mu \nu} + \frac{\phi_k}{16 \pi^2 f} G_{\mu \nu} {\tilde G}^{\mu \nu} \ .
\end{equation}
The term in the above equation involving only the scalar zero mode reads\footnote{Notice that since $\mathbb{M}^2$ is a real symmetric matrix (therefore it can be diagonalized by an orthogonal matrix), the scalar field
of the $j$-th site may be written in terms of mass eigenstates as $\phi_j = c_j \phi_{(0)} + ...$, where the dots denote strictly massive modes.}
\begin{equation}
	\mathcal{L}_4 \supset \frac{c_k \phi_{(0)}}{16 \pi^2 f} G_{\mu \nu} {\tilde G}^{\mu \nu}
				\equiv  \frac{\phi_{(0)}}{16 \pi^2 f_0} G_{\mu \nu} {\tilde G}^{\mu \nu} \ ,
\end{equation}
where we have defined an effective axion coupling scale
\begin{equation}
	f_0 = \frac{f}{c_k} = \frac{f q^k}{c_0} \simeq q^k f = q^k M_{Pl} \left( \frac{f}{M_{Pl}} \right) \ .
\label{eq:scalarf0}
\end{equation}
(We have restricted ourselves to the case of equal charges, $q_j \equiv q > 1$, for illustration.)
An effective axion coupling that is hierarchically larger than the symmetry breaking scale $f$ is dynamically generated
if the gauge theory is coupled to one of the scalar fields towards the end of the array of sites.

The clockwork mechanism for scalars then allows for exponentially different effective axion couplings depending on where the
gauge theory is localized, as a result of the symmetry-localization of the massless scalar field along the lattice,
and in the absence of site-dependent hierarchies in the decay constants of the $N+1$ axions in the unbroken phase.
In particular, for two non-abelian gauge theories localized on opposite sites,
but otherwise identical (with the same gauge coupling, and therefore the same physical properties like their confinement scales),
the clockwork mechanism leads to a hierarchy of effective axion couplings:
\begin{equation}
	\frac{f_{0,k=0}}{f_{0,k=N}} = q^{-N} \ll 1 \ .
\end{equation}

Finally, notice that $f_0$ can be super-Planckian in a natural fashion, in the sense that it is achieved with parametrically few lattice sites,
each of which may have a sub-Planckian symmetry breaking scale $f$.

\subsection{Abelian vector clockwork}
\label{sec:discreteVector}

In analogy with the scalar mechanism described in section~\ref{sec:discreteScalar},
the abelian vector clockwork \cite{Saraswat:2016eaz} consists of $N+1$ $U(1)$ gauge theories, each with its own gauge coupling $g_j$,
together with $N$ charge and mass-squared parameters, $q_j$ and $v_j^2$ ($j=0,...,N-1$), such that the lagrangian of the vector sector is given by
\begin{equation}
	\mathcal{L}_4 = - \sum_{j=0}^N \frac{1}{4 g_j^2} F_{j \mu \nu}^2 - \frac{1}{2} \sum_{j=0}^{N-1} v_j^2 (A_{j \mu} - q_j A_{j+1 \mu} )^2 \ .
\label{eq:L4quiver}
\end{equation}
The mass terms have the same form as those in eq.(\ref{eq:L4scalarDiscrete}) for the scalar case, and, as before,
may be regarded as arising from the vev's of $N$ scalar fields $\Phi_j$ ($j=0,...,N-1$)
with charges $+1$ and $-q_j$ under $U(1)_j$ and $U(1)_{j+1}$ respectively. As a result, $N$ of the $N+1$ abelian gauge symmetries are broken spontaneously, with a single unbroken $U(1)$ factor remaining.
Eq.(\ref{eq:L4quiver}) then corresponds to the effective lagrangian describing the vector sector, in unitary gauge.
The terms involving the only massless vector that remains in the spectrum are given by the substitutions $A_{j \mu} = c_j A_{\mu (0)} + \ldots$,
with $c_j$ as in eq.(\ref{eq:discreteScalarProfile}) and the dots denoting strictly massive modes,
yielding an effective gauge coupling
\begin{equation}
	\frac{1}{g_{(0)}^2} = \sum_{j=0}^N \frac{c_j^2}{g_j^2} \simeq \frac{c_0^2}{g^2}  \ ,
\label{eq:L4vectorGaugeCoupling}
\end{equation}
where in the last step we have assumed $q_j \equiv q > 1$ and $g_j \equiv g$ $\forall \ j$ for simplicity.\footnote{Strictly speaking,
in the gauge $U(1)$ case we consider here the coefficients $c_j$ are equal to those in eq.(\ref{eq:discreteScalarProfile}) for $j \geq 1$ with $c_0 = q^N$,
so that charge quantization in the $N$-th site in units of $g$ corresponds to charge quantization of the unbroken gauge theory in units of $g_{(0)} \simeq g q^{-N}$.}

If we now consider a scalar field $\varphi$ with charge $Q_{\varphi}$ under the $U(1)_k$ gauge group,
then its kinetic term reads
\begin{equation}
	\mathcal{L}_4 \supset - | ( \partial_\mu + i Q_{\varphi} A_{k \mu} ) \varphi |^2 \simeq - | ( \partial_\mu + i Q_{\varphi} c_0 q^{-k} A_{(0) \mu} + ... ) \varphi |^2\ ,
\label{eq:vectorMatterCoupling}
\end{equation}
where the dots denote strictly massive modes, and in the second equality we have again considered the case of $q_j \equiv q > 1$.
The effective coupling strength between $\varphi$ and the massless vector is then given by $\sim g_{(0)} Q_{\varphi} c_0 q^{-k} \simeq g Q_{\varphi} q^{-k}$.
In particular, for two scalar fields, $\varphi_0$ and $\varphi_N$, charged under the gauge groups at opposite sites with the same charge $Q_\varphi$,
the clockwork mechanism leads to an effective hierarchy of charges under the unbroken gauge group:
\begin{equation}
	\frac{Q_{0, k=N}}{Q_{0, k=0}} = q^{-N} \ll 1 \ .
\end{equation}
As in the scalar case, the exponential difference in effective couplings arises as a consequence of the symmetry-localization of the massless vector along the lattice,
and in the absence of site-dependent hierarchies in the gauge couplings of the $N+1$ vectors in the unbroken phase.

\subsection{(No) Non-abelian vector clockwork}
\label{sec:discreteNonAbelian}

The difficulties for constructing a non-abelian version of the discrete clockwork mechanism become apparent after having reviewed the abelian case.
By analogy, we might choose the $N$ scalar link fields, responsible for spontaneously breaking the non-abelian $G^{N+1}$ group down to $G$, to transform under different representations of adjacent gauge groups. However, as we show below, such a symmetry breaking pattern would not leave a single non-abelian symmetry group intact (the $N$ vev's would break all $N+1$ copies of $G$). The only viable lagrangian, which retains a $G$ symmetry after the link fields acquire vev's, has link fields transforming as bifundamentals, in which case it is clear that no clockworking can be generated, as this would be analogous to the abelian case discussed in section~\ref{sec:discreteVector} with all $q_j = 1$.

To illustrate this situation, consider $N+1$ copies of a non-abelian gauge group $SU(n)$, and $N$ scalar fields $\Phi_j$ ($j=0,...,N-1$)
transforming as bifundamentals under $SU(n)_j$ and $SU(n)_{j+1}$.
After spontaneous symmetry breaking of $N$ of the $N+1$ $SU(n)$ gauge symmetries due to the non-zero vev's of the scalar fields,
the effective lagrangian of the vector sector, in unitary gauge, is that of eq.(\ref{eq:L4quiver})
after setting $q_j \equiv 1$, and with the obvious replacements $A_{j \mu} \rightarrow A^a_{j \mu}$ and $F_{j \mu \nu} \rightarrow F^a_{j \mu \nu}$.
The massless vector lagrangian is then obtained by the substitutions $A^a_{j \mu} = A^a_{\mu (0)} + \ldots$, and the effective gauge coupling
of the unbroken non-abelian gauge theory is given by
\begin{equation}
	\frac{1}{g_{(0)}^2} = \sum_{j=0}^{N} \frac{1}{g_j^2} \ .
\label{eq:profile4DnonAbelian}
\end{equation}

Consider now a scalar field $\varphi$ transforming under a representation $\mathcal{R}$ of the gauge group $SU(n)_k$.
Its kinetic term reads
\begin{equation}
	\mathcal{L}_4 \supset - | ( \partial_\mu + i A^a_{k \mu} T^a_{\mathcal{R}} ) \varphi |^2 = - | ( \partial_\mu + i A^a_{(0) \mu} T^a_{\mathcal{R}} + ... ) \varphi |^2 \ ,
\end{equation} 
where $T^a_{\mathcal{R}}$ are the generators of $SU(n)$ in the appropriate representation, and the dots denote strictly massive modes.
The field $\varphi$ then transforms under representation $\mathcal{R}$ of the unbroken $SU(n)$ factor, with an effective gauge coupling $g_{(0)}$
independent of the position of the $k$-th site.\footnote{This is a hardly surprising result, for our construction is manifestly gauge invariant,
and a massless state with different effective gauge couplings to different matter fields would violate gauge invariance explicitly.}

Moreover, notice from eq.(\ref{eq:profile4DnonAbelian}) that it is not possible to generate a parametrically small effective gauge coupling
in a natural fashion. In particular, eq.(\ref{eq:profile4DnonAbelian}) has two ineffective limits. One, we may set all $g_j=g$, such that
$g_{(0)} \simeq g / \sqrt{N}$, and so an unnaturally large number of sites $N$ would be required to
generate a meaningful hierarchy between $g_{(0)}$ and $g_j$. Two, the individual $g_j$ may be of parametrically different sizes,
the smallest of which determines the size of $g_{(0)} \sim \min_{j} g_j$.

We can be more general, and prove that the lack of symmetry-localization of the massless vector mode along the different sites is in fact a requirement
if its mass is to be protected by gauge invariance.\footnote{Above, we have only shown that non-abelian clockwork cannot arise if the $N$ scalar fields
transform as bifundamentals. However, one could ask whether a more complicated construction (for instance, the case in which each $\Phi_j$ transforms
under inequivalent representations of contiguous gauge groups, or a construction that is not restricted to nearest neighbor interactions)
could lead to consistent non-abelian clockwork.}
To illustrate this, consider the case in which the vector field on every site is given by $A^a_{j \mu} = c_j A^a_{(0) \mu} + ...$, with the dots denoting massive modes
as usual, and let's remain agnostic about the dynamical origin of the coefficients $c_j$.
The kinetic terms of the $N+1$ non-abelian gauge theories read
\begin{equation}
\begin{split}
	\mathcal{L}_{4, {\rm kin}} & = - \sum_{j=0}^N \frac{1}{4 g_j^2} \left( F^{a}_{j \mu \nu} \right)^2 \\
			& = - \sum_{j=0}^N \frac{1}{g_j^2} \left\{ \frac{1}{4} \left( \hat{F}^{a}_{j \mu \nu} \right)^2 + 
			f^{abc} \partial_\mu A_{j \nu}^a A^{b \mu}_j A^{c \nu}_j + \frac{1}{4} f^{abc} f^{ars} A^b_{j \mu} A^c_{j \nu} A^{r \mu}_j A^{s \nu}_j \right\} \ ,
\end{split}
\label{eq:nonabelianKin}
\end{equation}
where $\hat{F}^a_{j \mu \nu} \equiv \partial_\mu A_{j \nu}^a - \partial_\nu A_{j \mu}^a$.
Substituting $A^a_{j \mu} = c_j A^a_{(0) \mu} + ...$, the terms in eq.(\ref{eq:nonabelianKin}) involving the massless mode $A^a_{(0) \mu}$ only read
\begin{equation}
\begin{split}
	\mathcal{L}_{4, {\rm kin}} \supset & - \frac{1}{4} \left( \sum_{j=0}^N \frac{c_j^2}{g_j^2} \right) \hat{F}^{a \ 2}_{(0) \mu \nu}
			- \left( \sum_{j=0}^N \frac{c_j^3}{g_j^2} \right) f^{abc} \partial_\mu A_{(0) \nu}^a A^{b \mu}_{(0)} A^{c \nu}_{(0)} \\
			& - \frac{1}{4} \left( \sum_{j=0}^N \frac{c_j^4}{g_j^2} \right) f^{abc} f^{ars} A^b_{(0) \mu} A^c_{(0) \nu} A^{r \mu}_{(0)} A^{s \nu}_{(0)} \ .
\end{split}
\label{eq:nonabelianKin0mode}
\end{equation}
Gauge invariance of the massless mode lagrangian requires all three sums in the equation above be equal,\footnote{Gauge invariance
requires both the terms in eq.(\ref{eq:nonabelianKin0mode}), and interaction terms between the massless mode and massive modes
(omitted from eq.(\ref{eq:nonabelianKin0mode})) be gauge invariant independently.
However, focusing on the terms in eq.(\ref{eq:nonabelianKin0mode}) will be sufficient to prove that it is not possible to build non-abelian clockwork.}
and they define the effective gauge
coupling of the unbroken theory, i.e.~
\begin{equation}
	\frac{1}{g_{(0)}^2} \equiv \sum_{j=0}^N \frac{c_j^2}{g_j^2} = \sum_{j=0}^N \frac{c_j^3}{g_j^2} = \sum_{j=0}^N \frac{c_j^4}{g_j^2} \ .
\label{eq:gaugeInvNonAbelian}
\end{equation}
The above equalities are only satisfied if $c_j \in \{0,1\}$ $\forall j$, and
the terms in eq.(\ref{eq:nonabelianKin0mode}) are then manifestly invariant under infinitesimal gauge transformations of the usual form
$A^a_{(0) \mu} \rightarrow A_{(0) \mu}^a + \partial_\mu \alpha^a - f^{abc} \alpha^b A_{(0) \mu}^c$.

This general argument addresses, in particular, the case in which the scalar fields $\Phi_j$ are chosen to transform under inequivalent
representations of the gauge groups at sites $j$ and $j+1$, as well as more intricate constructions in which the $\Phi_j$ are chosen
to transform non-trivially under non-contiguous gauge groups.
Either way, the resulting effective lagrangian describing the vector sector will not have a clockworked
non-abelian gauge boson.

Thus, although it is possible to build constructions leading to $c_j \notin \{ 0, 1 \}$,
and in which the lowest lying vector mode is massless at tree-level
(e.g.~by writing a mass term for the non-abelian gauge sector as in eq.(\ref{eq:L4quiver}) with $q_j>1$),
the masslessness of this mode will not be protected by gauge invariance.
We can therefore conclude that a meaningful clockwork mechanism is impossible to engineer in the context of a non-abelian gauge theory.
As we discuss next in section~\ref{sec:discreteGraviton}, this statement straightforwardly generalizes to the graviton case --
an unsurprising result, for gravity is a non-abelian theory itself.

\subsection{(No) Graviton clockwork}
\label{sec:discreteGraviton}

After having discussed the scalar and vector cases, one could wonder whether a spin-2 version of the clockwork
mechanism may be consistently built.
As before, the starting point would consist of $N+1$ sites, each of them with its own metric $g_{j \mu \nu}$ and general
coordinate invariance symmetry ${\rm GC}_j$.
Allowing for gravitational interactions of varying strength on each site, the Einstein-Hilbert part of the lagrangian simply reads
\begin{equation}
	\mathcal{L}_{4, {\rm EH}} = \sum_{j=0}^N \frac{M_j^2}{2} \sqrt{|g_j|} R_j \ ,
\label{eq:kinTermGrav}
\end{equation}
where $R_j$ is the Ricci scalar corresponding to the metric $g_{j \mu \nu}$, and $M_j$ the reduced Planck mass at site $j$.
Eq.(\ref{eq:kinTermGrav}) is manifestly invariant under all $N+1$ copies of ${\rm GC}_j$.
If we expand the metric on every site as a perturbation around flat space, i.e.~$g_{j \mu \nu} = \eta_{\mu \nu} + h_{j \mu \nu}$,
then the expansion of eq.(\ref{eq:kinTermGrav}) up to $\mathcal{O}(h_j^2)$ takes the familiar form
\begin{equation}
	\mathcal{L}_{4, {\rm EH}} = \sum_{j=0}^N \frac{M_j^2}{2} \left\{ - \frac{1}{4} (\partial_\mu h_{j \rho \sigma})^2 + \frac{1}{4} (\partial_\mu h_j)^2
				+ \frac{1}{2} (\partial_\mu h_j^{\mu \nu})^2 - \frac{1}{2} \partial^\mu h_j \partial^\nu h_{j \mu \nu} + ... \right\} \ ,
\label{eq:kinTermGravLin}
\end{equation}
where $h_j \equiv \eta_{\mu \nu} h_j^{\mu \nu}$.

Subtleties arise when trying to write a mass term that would render $N$ of the gravitons massive in a way that allows for
the full general coordinate invariance of the theory to be restored at some high scale.
This was thoroughly explored in~\cite{ArkaniHamed:2002sp}, where it is argued that this may be achieved by introducing $N$ `link'
fields $Y_j^\mu$ ($j=0,...,N-1$), which transform non-trivially under ${\rm GC}_j$ and ${\rm GC}_{j+1}$,
in complete analogy with the scalar fields $\Phi_j$ introduced in the vector case.
As discussed in~\cite{ArkaniHamed:2002sp}, each field $Y_j^\mu$ corresponds to a map between a set of coordinates $x^\mu_j$ at site $j$
and coordinates  $Y^\mu_j (x_j)$ at site $j+1$, and defines a pullback map from site $j+1$ to site $j$.
For instance, using this map we can pullback the metric $g_{j+1 \mu \nu}$, which is defined at site $j+1$ and transforms non-trivially
under ${\rm GC}_{j+1}$, to find an object
\begin{equation}
	G_{j \mu \nu} (x_j) \equiv \frac{\partial Y^\alpha_j}{\partial x_j^\mu} \frac{\partial Y^\beta_j}{\partial x_j^\nu} g_{j+1 \alpha \beta} ( Y_j (x_j) ) \ ,
\end{equation}
which is now defined at site $j$, and transforms as a metric under ${\rm GC}_{j}$.
In particular, it is now possible to add a term to the lagrangian that respects the full general coordinate invariance of the theory, of the form~\cite{ArkaniHamed:2002sp}
\begin{equation}
	\mathcal{L}_{4} \supset \frac{1}{2} \sum_{j=0}^{N-1} \sqrt{|g_j|} \frac{M_j^2 m_j^2}{4}
				(g_{j \mu \nu} - G_{j \mu \nu}) (g_{j \alpha \beta} - G_{j \alpha \beta}) (g_j^{\mu \nu} g_j^{\alpha \beta} - g_j^{\mu \alpha} g_j^{\nu \beta}) \ ,
\label{eq:massTermGrav}
\end{equation}
where the mass parameters $m_j$ will set the mass scale of the $N$ massive graviton excitations, and are analogous to the mass parameters introduced in the
scalar and vector cases.

Since we are interested in expanding the metric on every site around the same flat space background,
unitary gauge corresponds to $Y_j^\mu = x_j^\mu$ $\forall \ j=0,...,N-1$.\footnote{This is not necessarily the case in general,
but it holds if we are expanding around flat space in each site,
since in this case the pullback of the background metric acting at site $j+1$ must be equal to the background metric acting at site $j$.}
In this gauge, the terms in eq.(\ref{eq:massTermGrav}) that are quadratic in the perturbation lead to a mass term
\begin{equation}
	\mathcal{L}_{4, {\rm mass}} = \frac{1}{2} \sum_{j=0}^{N-1} \frac{M_j^2 m_j^2}{4}
				\left\{ ( h_j - h_{j+1} )^2 - ( h_{j \mu \nu} - h_{j+1 \mu \nu} )^2 \right\} \ .
\label{eq:massTermGravLin}
\end{equation}
As in the non-abelian case, the massless graviton lagrangian can be obtained by the substitutions $h_{j \mu \nu} = h_{(0) \mu \nu} + ...$,
where the dots denote strictly massive states,
and eq.(\ref{eq:kinTermGravLin}) then defines an effective 4D Planck scale
\begin{equation}
	M_{(0)} = \left( \sum_{j=0}^{N} M_j^2 \right)^{1/2} \ .
\label{eq:effectiveMPl}
\end{equation}
This expression clearly illustrates how an effective scale $M_{(0)}$ much larger than the fundamental scale $M_j$ of
the individual sites is not possible to engineer in this context.
In particular, in the simplest case $M_j \equiv M$ $\forall \ j$, $M_{(0)} \simeq M \sqrt{N}$,
and so a large hierarchy between $M_{(0)}$ and $M$ would require an even larger value of $N$,
frustrating any attempt to build a solution to the electroweak hierarchy problem in a natural fashion.

If we now consider a stress energy tensor defined on the $k$-th site, its leading coupling to the metric perturbation is of the form
\begin{equation}
	\mathcal{L}_4 \propto h_{k \mu \nu} T^{\mu \nu} = h_{(0) \mu \nu} T^{\mu \nu} + ... \ ,
\label{eq:gravitonMatter}
\end{equation}
where the dots denote strictly massive graviton modes.
We see how the massless graviton couples with the same strength to a given stress-energy tensor,
independently of the position of the site in which $T^{\mu \nu}$ is defined, in keeping with the Equivalence Principle.

As in the non-abelian case of section~\ref{sec:discreteNonAbelian}, we can be more general and prove that the flatness of the massless graviton mode
across the different sites is again a requirement if its mass is to be protected by diffeomorphism invariance.\footnote{So far, we have only shown that
a term like that of eq.(\ref{eq:massTermGrav}) does not lead to an asymmetrically distributed massless graviton.
However, one could ask whether a more complicated version of eq.(\ref{eq:massTermGrav}) could lead, at quadratic order,
to an effective mass term like that in eq.(\ref{eq:massTermGravLin}) but with asymmetric couplings in front of the $h_j$ and $h_{j+1}$ terms.}
In order to do this, it is crucial to consider terms in the expansion of eq.(\ref{eq:kinTermGrav}) that involve higher-order
terms in the metric perturbation.
Schematically, such an expansion has the form
\begin{equation}
	\mathcal{L}_{4, {\rm EH}} \sim \sum_{j=0}^N M_j^2 \left\{ \partial^2 h_j^2 + \sum_n \partial^2 h_j^{2+n} \right\} \ .
\end{equation}
Now, if we allow ourselves to write $h_{j \mu \nu} = c_j h_{(0) \mu \nu} + ...$, without prejudice about the origin of the $c_j$ coefficients,
then the previous equation reads
\begin{equation}
	\mathcal{L}_{4, {\rm EH}} \sim \left( \sum_{j=0}^N M_j^2 c_j^2 \right) \partial^2 h_{(0)}^2 + \sum_n  \left( \sum_{j=0}^N M_j^2 c_j^{2+n} \right) \partial^2 h_{(0)}^{2+n} + ... \ ,
\label{eq:gravExpansion0mode}
\end{equation}
where the dots denote terms involving massive graviton modes only, but also interaction terms between the massless and massive gravitons.
As in the non-abelian case, that the terms in the effective lagrangian involving the massless graviton be diffeomorphism invariant
requires that all the sums in the equation above be equal,\footnote{Again, we emphasize that diffeomorphism invariance
requires both the terms explicitly written in eq.(\ref{eq:gravExpansion0mode}), and interaction terms between massless and massive modes
be invariant independently. However, focusing on the terms in eq.(\ref{eq:gravExpansion0mode}) will be enough to rule out the possibility
of building a clockwork graviton.} and define an effective Planck scale, i.e.~
\begin{equation}
	M_{(0)}^2 \equiv \sum_{j=0}^N M_j^2 c_j^2 = \sum_{j=0}^N M_j^2 c_j^{2+n} \quad \forall \ n \geq 1 \ .
\end{equation}
As in section \ref{sec:discreteNonAbelian}, these equalities are only satisfied for $c_j \in \{ 0 , 1\}$.
The terms in eq.(\ref{eq:gravExpansion0mode}) are then invariant under infinitesimal diffeomorphism transformations
of the usual form
\begin{equation}
	h_{(0) \mu \nu} \rightarrow h_{(0) \mu \nu} + \partial_\mu \epsilon_\nu + \partial_\nu \epsilon_\mu +
		f^{\rho \sigma}_{\mu \nu} \partial_\rho \epsilon^\alpha h_{(0) \alpha \sigma} + \epsilon^\alpha \partial_\alpha h_{\mu \nu} \ ,
\label{eq:diffeom}
\end{equation}
where $f^{\rho \sigma}_{\mu \nu} \equiv \delta^\rho_\mu \delta^\sigma_\nu + \delta^\rho_\nu \delta^\sigma_\mu$.\footnote{We remind the reader that eq.(\ref{eq:diffeom})
is the way in which the metric perturbation $h_{\mu \nu}$ changes under an infinitesimal diffeomorphism transformation, which in a coordinate basis is
given by $Y^\mu = x^\mu + \epsilon^\mu$, regardless of the size of $h_{\mu \nu}$. The last term in eq.(\ref{eq:diffeom}) captures the non-abelian
nature of gravity, and must be taken into account if we want to assess whether the masslessness of the graviton is indeed symmetry-protected.}
Hence, any construction that leads to $c_j \notin \{ 0 , 1\}$ will feature a lowest-lying graviton excitation whose
mass is not protected by diffeomorphism invariance, even if it is engineered to be massless at tree-level.

In analogy with the results of section~\ref{sec:discreteNonAbelian} for non-abelian gauge fields, we conclude that it is not possible
to build a 4D effective theory in which a massless spin-2 particle is symmetry-localized
and, at the same time, retains diffeomorphism invariance.
Moreover, in the absence of exponential hierarchies among the values of the different scales $M_j$,
the effective Planck scale only depends on the number of sites as $\sim \sqrt{N}$.
Consequently, it is apparent that there is no such thing as a clockwork graviton, and, by extension, no such thing as a clockwork solution to the hierarchy problem.
(In appendix~\ref{app:graviton} we explicitly show how a clockwork graviton does \emph{not} arise
when deconstructing a gravitational extra dimension.)

As an aside, we note that it is sometimes common, and convenient, to rescale the metric perturbation as $h_{j \mu \nu} \rightarrow 2 h_{j \mu \nu} / M_j$,
so that the kinetic terms in eq.(\ref{eq:kinTermGravLin}) are canonical.
In this rescaled basis, eq.(\ref{eq:massTermGravLin}) now reads
\begin{equation}
	\mathcal{L}_{4, {\rm mass}} = \frac{1}{2} \sum_{j=0}^{N-1} m_j^2
				\left\{ \left( h_j - \frac{M_j}{M_{j+1}} h_{j+1} \right)^2 - \left( h_{j \mu \nu} - \frac{M_j}{M_{j+1}} h_{j+1 \mu \nu} \right)^2 \right\} \ ,
\label{eq:massTermGravLinNorm}
\end{equation}
and the massless graviton mode is just given by $h_{(0) \mu \nu} = \sum_{j=0}^N (M_j / M_{(0)}) h_{j \mu \nu}$.\footnote{Notice
eq.(\ref{eq:massTermGravLinNorm}) has the form of eq.(2.35) in~\cite{CW}, but with the extra necessary condition $q_j = M_j / M_{j+1}$, i.e.~non-unit $q$'s are only a consistent choice in the presence of an exponential distribution of Planck scales.}
The graviton coupling to matter in eq.(\ref{eq:gravitonMatter}) is now
\begin{equation}
	\mathcal{L}_4 \propto \frac{h_{k \mu \nu}}{M_k} T^{\mu \nu} = \frac{h_{(0) \mu \nu}}{M_{(0)}} T^{\mu \nu} + ... \ ,
\end{equation}
which again makes it explicit how the strength of gravitational interactions between matter and the massless graviton mode
is just set by $M_{(0)}$, as given in eq.(\ref{eq:effectiveMPl}).

\subsection{When does clockwork not work?}\label{sec:symmetries}

The results of the previous sections can also be understood clearly from the perspective of the unbroken clockwork symmetry, both in the low-energy effective theory and in possible UV completions.
For simplicity we will focus here on abelian vector clockwork, for which the role of the clockwork symmetry is particularly clear, though the conclusions apply equally well to all spins, and clarify the cases in which meaningful clockwork is possible. 

The abelian vector clockwork of eq.(\ref{eq:L4quiver}) arises from a UV theory of $N+1$ $U(1)$ gauge bosons connected by $N$ link scalar fields $\Phi_j$ via
\begin{equation} \label{eq:goodvectorclockwork}
\mathcal{L}_4 = -\sum_{j=0}^N \frac{1}{4 g_j^2} F^2_{j \mu \nu} - \sum_{j=0}^{N-1} |D_\mu \Phi_j |^2 + \dots \ ,
\end{equation}
where $D_\mu \Phi_j \equiv \left[ \partial_\mu + i \left(A_{j \mu} - q_j A_{j+1 \mu} \right) \right] \Phi_j$ and the dots denote, e.g., potentials for the $\Phi_j$. For simplicity, we will focus on the case of equal charges, couplings, and symmetry breaking scales, but our conclusions hold for any theory in which there are no large hierarchies. If the $\Phi_j$ acquire vacuum expectation values $\langle | \Phi_j |^2 \rangle = f^2 / 2 $, this results in a clockwork mass matrix for canonically normalized gauge fields of the form
\begin{equation} \label{eq:vectormassmatrix}
- \sum_{j=0}^{N-1} \frac{g^2 f^2}{2} \left(A_{j \mu} - q A_{j+1 \mu} \right)^2 \ .
\end{equation}
In order to probe the unbroken clockwork symmetry, we introduce a matter field $\varphi$ charged under the $U(1)$ gauge group of site $k$ with charge $Q_\varphi$. The clockwork gauge symmetry preserved by eq.(\ref{eq:vectormassmatrix}) corresponds to $A_{j \mu} \rightarrow A_{j \mu} + \partial_\mu \alpha (x)/q^j$ $\forall j$. Under such a gauge transformation, $\varphi \rightarrow e^{i \alpha Q_\varphi / q^k} \varphi$, which is naturally interpreted as a small and site-dependent charge $Q_\varphi / q^k$ under the unbroken $U(1)$. This makes clear the sense in which the site-dependent charges found in section \ref{sec:discreteVector} are a direct probe of the asymmetric distribution of the clockwork symmetry among different sites.

Considering clockwork from the perspective of the unbroken symmetry also makes apparent the sense in which theories with the mass matrix eq.(\ref{eq:vectormassmatrix}) may {\it fail} to generate clockwork. In particular, the clockwork theory of eq.(\ref{eq:goodvectorclockwork}) without any large hierarchies of couplings, charges, and scales (``Theory A'') is not the only way of generating the mass matrix in eq.(\ref{eq:vectormassmatrix}). An identical mass matrix arises in a theory (``Theory B'') of $N+1$ $U(1)$ gauge bosons with $N$ bifundamental scalars $\Phi_j$, likewise described by eq.(\ref{eq:goodvectorclockwork}), in which the $\Phi_j$ carry opposite charges under adjacent groups ($q_j = 1$), the $g_j$ are unequal and satisfy $g_{j+1}/g_j = q$, and the vacuum expectation values $v_j$ of the scalars $\Phi_j$ satisfy $g_j^2 v_j^2 = g^2 f^2$. Notably, there is an exponential hierarchy between the couplings and vev's at either end of the Theory B quiver, $g_N /g_0 = v_0/v_N = q^N$. Such a theory likewise preserves a $U(1)$ symmetry, but one that is symmetrically distributed among sites and exhibits no clockwork phenomena. Given a probe field $\varphi$ of charge $Q_\varphi$ on the site $k$, a gauge transformation of the unbroken $U(1)$ symmetry induces a rotation of the probe field by $e^{i \alpha Q_\varphi}$, independent of the position of the site. This universality is born out by diagonalizing the mass matrix and studying the couplings of the massless gauge field: the zero mode is $\propto \sum_{j=0}^N g_j^{-1} A_{j \mu}$, and therefore couples universally to matter fields on different sites. This theory does not clockwork, though it shares the mass matrix of eq.(\ref{eq:vectormassmatrix}) with a theory that {\it does}. 

One might object that Theory A and Theory B are actually the same theory, related by rescaling the gauge kinetic terms and the charges of both the link fields $\Phi_j$ and the probe fields $\varphi$ in Theory B to match those of Theory A, so that there is no invariant distinction between the two. This is certainly true if the gauge group at each site is taken to be $\mathbb{R}$ rather than $U(1)$, but in this case there is no notion of natural charge assignments and clockwork is uninteresting to begin with. Rather, an invariant distinction exists when additional criteria restrict the gauge groups to genuine $U(1)$s and fully specify the spectrum of electric and magnetic charges, as is the case in a theory of quantum gravity.

In a theory of quantum gravity (including all known examples in string theory), all continuous gauge groups are compact and satisfy the Completeness Hypothesis \cite{Banks:2010zn}, namely that every electric and magnetic charge allowed by Dirac quantization is present in the spectrum. In this case, Theory A possesses a spectrum of states at each site carrying all possible electric charges $n \in \mathbb{Z}$ (in units of $g_j = g$) and all possible magnetic charges $2 \pi n / g_j$. Theory B possesses a similarly complete spectrum, but with respect to the exponentially varying $g_j$. Rescaling the charges and couplings of Theory B to match those of Theory A leads to a gap in the spectrum of electric and magnetic charges at each site, in conflict with the Completeness Hypothesis. Equivalently, the spectrum of states charged under the unbroken $U(1)$ differs between the two theories. In Theory A, the number of states of charge $Q \in \nn$, in units of the effective coupling of the massless $U(1)$, is the largest $i \leq N+1$ for which $q^i$ divides $Q$. However, in Theory B, there are simply $N+1$ states of any given charge under the unbroken $U(1)$, which attests to the diagonal nature of the symmetry breaking in this latter case. For instance, in Theory A there is only one state of unit electric charge in units of the effective coupling of the massless $U(1)$, while in Theory B there are $N+1$ such states with unit electric charge under the unbroken $U(1)$. Thus Theory A and Theory B are genuinely distinct theories, with distinct physical observables, and only the former exhibits clockwork phenomena. 

The distinction between the two theories is not merely academic, but is essential for generating natural exponential hierarchies in a theory of quantum gravity. For example, Theory A can satisfy the magnetic form of the Weak Gravity Conjecture (WGC) in the UV, but upon higgsing gives rise to an effective theory for the massless $U(1)$ that exponentially violates the magnetic WGC \cite{Saraswat:2016eaz}. This is a precise sense in which the clockwork mechanism is a useful generator of natural exponential hierarchies. In contrast, if Theory B satisfies the magnetic WGC in the UV, then the effective theory of the massless $U(1)$ also trivially satisfies the magnetic WGC. Theory B generates no useful exponential hierarchies -- rather, it requires them as inputs.

Aside from quantum gravity arguments, discerning whether an abelian gauge theory `clockworks' or not requires making reference to a localized
lattice of charged states. The requirement that states with the same integer charge on different sites have (exponentially) different charges under the unbroken gauge theory singles out models with symmetry-localized zero modes as the only ones that can exhibit clockwork dynamics.

As we will see, the distinction between Theory A and Theory B becomes important when attempting to identify the continuum equivalent of discrete clockwork in an extra dimension. One can always find a metric for which the Kaluza-Klein decomposition of a bulk field gives rise to the mass matrix in eq.(\ref{eq:vectormassmatrix}). But as we have argued, this alone is not enough for the continuum theory to generate clockwork. Whether the continuum theory provides a successful realization of clockwork depends on whether its discretization gives Theory A or Theory B. More precisely, continuum clockwork requires a compact 5D $U(1)$ gauge theory to lead, upon compactification, to a 4D effective gauge theory that is non-compact.

While we have focused on abelian vector clockwork, one would expect that identical arguments go through for abelian scalar clockwork whenever there exists a well-defined notion of an asymmetrically-distributed global symmetry (see \cite{ChangSub}). For example, in a UV completion of scalar clockwork, the roles of gauge transformations and probe charges in vector clockwork are played by global symmetry transformations and anomaly coefficients. The connection should become particularly transparent when one considers that all apparent global symmetries should originate as gauge symmetries in a theory of quantum gravity. 

Finally, the distinction between Theory A and Theory B makes clear why clockwork is an inherently abelian phenomenon. While one is free to choose the charges and couplings in a quiver theory with  abelian symmetry factors to obtain Theory A or Theory B, in a quiver theory with non-abelian symmetry groups the only option consistent with the symmetries is the non-abelian version of Theory B, as we will now see more rigorously.

\section{Group theory of clockwork}
\label{sec:GT}

In the discrete clockwork models of sections \ref{sec:discreteScalar}--\ref{sec:discreteVector}, a $U(1)^{N+1}$ symmetry is broken to a single $U(1)$. The massless mode transforms under, and has its mass protected by, the remaining $U(1)$ symmetry. Moreover, the remaining $U(1)$ effects unequal rotations at the $N+1$ sites of the original quiver, which results in the unequal couplings of the massless mode to matter localized at different sites.

Our initial attempts in sections \ref{sec:discreteNonAbelian}--\ref{sec:discreteGraviton} to construct something similar for other symmetry groups were unsuccessful. Which begs the question: for a general group $G$, could we ever design a pattern of symmetry breaking such that $G^{N+1} \to G$, and such that the remaining $G$ also acts unequally on the $N+1$ sites of the quiver? In this section we offer a heuristic argument which rules out any mechanism directly analogous to clockwork for a non-abelian Lie group $G$. In appendix \ref{app:group}, we consider a generic group $G$, and derive the conditions on $G$ under which such a subgroup's action on the lattice sites can be in any way asymmetric.

We recall that the kinetic terms of the discrete clockwork lagrangian (\ref{eq:L4scalarDiscrete}) are invariant under independent shifts of the $\phi_j \to \phi_j + \alpha_j$. The mass terms,
\begin{equation}
- \frac{1}{2} \sum_{j=0}^{N-1} m_j^2 (\phi_j - q_j \phi_{j+1})^2,
\end{equation}
break the independent shift symmetries such that, for a given shift $\phi_{0} \to \phi_{0} + \alpha$, the lagrangian is only invariant if $\phi_1 \to \phi_1 + \frac{\alpha}{q_0}$. By iteration, the corresponding shifts of the other $\phi_j$ at the rest of the lattice sites are also fixed.
If the generator of the shift symmetry at the $j$th lattice site is $T_j$, the generator of the unbroken shift symmetry
\begin{equation}
T_{(0)} \propto T_0 + \frac{1}{q_0} T_{1} + \ldots + \left( \prod_{k=0}^{N-1} \frac{1}{q_k} \right) T_N,
\end{equation}
which in turn generates the massless mode's hierarchy of couplings to different sites seen in (\ref{eq:discreteScalarProfile}).

Suppose we construct a similarly weighted set of generators in the non-abelian case. Write $T_j^a$ as the generators of the symmetry group $G$ on the $j$th lattice site, where $a=1,\ldots,\dim G$ is an adjoint index. Define the Lie bracket as $[T_i^a,T_j^b] = f^{abc} T_j^c \delta_{ij}$, summing over repeated adjoint indices. Assume
\begin{equation}
T_{(0)}^a = \sum_{j=0}^{N} a_j T_j^a.
\label{eq:weightSubalgebra}
\end{equation}
Requiring that $[T_{(0)}^a,T_{(0)}^b] = f^{abc} T_{(0)}^c$, we find
\begin{equation}
\sum_{j=0}^{N} f^{abc} a_j^2 T^c_j = \sum_{j=0}^{N} f^{abc} a_j T^c_j.
\end{equation}
As at least one structure constant $f^{abc} \neq 0$, we conclude
\begin{equation}
a_j^2 = a_j, \forall j
\implies a_j \in \{0,1\}, \forall j .
\end{equation}
Thus, of the lattice sites which transform at all under the unbroken generators, they must all be shifted equally.

Importantly, for models whose fields form a non-abelian Lie algebra, such as gauge theories, non-linear sigma models or gravitons (and whose mass is therefore protected by the resulting symmetry), massless modes must therefore couple universally to different lattice sites.\footnote{One might wonder whether this is the whole story, as one can choose the normalization of the Lie algebra valued forms at each site of the quiver. For instance, one might rescale the gauge field $A_j$ such that it transforms as $A_j \to U A_j U^{-1} + \frac{1}{g_j} U \der U^{-1}, U \in G_j$ for $j$ dependent $g_j$. However, the observable physics will ultimately be invariant under such field rescalings. Namely, shifts of the massless mode $A_{(0)}$ must always be \emph{proportional} to the site independent shifts of the corresponding subalgebra $U \der U^{-1}, U \in G$.} We thus conclude that clockwork, insofar that it relies on the construction of a weighted subalgebra of the form (\ref{eq:weightSubalgebra}), is a strictly abelian phenomenon.
 
The above is suggestive of the more general group theory result of appendix \ref{app:group}; unless $G$ has a non-trivial normal subgroup $K$ whose corresponding quotient group $G/K$ is also a subgroup of $G$, it is impossible to construct any subgroup which acts unevenly on different lattice sites (let alone a subgroup whose algebra is a weighted combination of the sites' generators as mooted above). To wit, for any subgroup $G < G^{N+1}$, the action of any subgroup element $g \in G$ on an element $g_j \in G_j$ (the $j$th group in the direct product) is given by $\varphi_j(g) g_j$. $\varphi_j: G \to G_j$ must be a group isomorphism or a map to the trivial group within $G_j$,
unless $G$ satisfies the aforementioned conditions.

\section{No clockwork from geometry}
\label{sec:noMasses}

After having rigorously established in sections \ref{sec:discrete} and \ref{sec:GT} that it is only possible to build consistent
clockwork models in the spin-0 and abelian spin-1 cases, we now set to answer the question of whether such discrete models
could arise from the deconstruction of 5D theories in which the corresponding bosonic fields propagate in a non-trivial background.
We find that the answer is negative: geometry alone cannot clockwork bosonic fields.
This statement is true in that neither the continuum theory nor its deconstruction exhibits position dependent couplings
as a consequence of a symmetry-localization of the scalar or vector massless modes.
In particular, we establish how, at best, it is possible to \emph{accommodate} the discrete clockwork models of sections~\ref{sec:discreteScalar} and \ref{sec:discreteVector}
as the deconstruction of 5D theories with conformally flat metrics, but then \emph{ad hoc} exponential hierarchies in the couplings between bulk
and brane fields need to be introduced in order for the deconstruction to match clockwork.  This is true in particular of 5D theories in linear dilaton backgrounds, as considered in~\cite{CW},
and in section \ref{sec:dilaton} we make it explicit how couplings involving the dilaton field do not change the above statements. In the language of section \ref{sec:symmetries}, linear dilaton backgrounds always give the unclockworked Theory B. This is particularly clear in the vector case, where continuum clockwork requires a non-compact (therefore $\mathbb{R}$) 4D effective abelian gauge symmetry arising from a compact 5D symmetry (i.e.~a genuine $U(1)$). As we emphasize in this section, geometry alone only allows compact higher-dimensional gauge theories to
generate compact 4D effective ones, therefore precluding any kind of clockwork dynamics.

We consider an extra dimension compactified on an $S^1 / \mathbb{Z}_2$ orbifold, with the fifth dimension parametrized by a coordinate $y$,
and with two end-of-the-world branes present at the orbifold fixed points ($y=0$ and $y=\pi R$).
We will focus on the case in which both the scalar and gauge fields are even under the orbifolding $\mathbb{Z}_2$ symmetry,
in order to allow for a massless state to be present in the spectrum of KK-modes.
In keeping with the notation introduced in \cite{CW}, we consider a background metric of the general form
\begin{equation}
	ds^2 = g_{M N} d x^M d x^N = X(y) d x_\mu d x^\mu + Y(y) dy^2 \ ,
\label{eq:metric}
\end{equation}
with $y \in [0, \pi R] $.
We consider a bulk scalar field coupled to a brane-localized non-abelian gauge theory in section~\ref{sec:scalarNoMasses},
and then discuss the case of a bulk $U(1)$ gauge theory in section~\ref{sec:vectorNoMasses}.

\subsection{Scalar case}
\label{sec:scalarNoMasses}

The action of a massless, non-interacting real scalar field propagating in a non-trivial background is given by
\begin{equation}
	S_{5, {\rm bulk}} = - \frac{1}{2} \int d^4 x d y \sqrt{|g|} g^{MN} \partial_M \phi \partial_N \phi \ .
\label{eq:S5scalar}
\end{equation}
In a background of the form given in eq.(\ref{eq:metric}), and after expanding the 5D scalar field $\phi$ as a sum over KK-modes as
$\phi = \sum_{n = 0}^\infty \chi_n (y) \phi^{(n)} (x)$, the equations of motion and boundary conditions for the different modes read
\begin{equation}
	\partial_y \left( \frac{X^2}{\sqrt{Y}} \partial_y \chi_n \right) + m_n^2 \chi_n X \sqrt{Y} = 0
\label{eq:scalarEOM}
\end{equation}
\begin{equation}
	\partial_y \chi_n = 0 \qquad {\rm at} \qquad y=0, \pi R \ ,
\label{eq:scalarBC}
\end{equation}
where $m_n^2$ corresponds to the mass-squared of the $n$-th KK-mode excitation.
In particular, a massless mode is present in the KK-spectrum, whose profile is a constant
$\chi_0 (y) = {\mathcal C}_0$.\footnote{For a canonically normalized scalar field ${\mathcal C}_0 = \left( \int_0^{\pi R} dy X(y) \sqrt{Y(y)} \right)^{-1/2}$.}

If we now consider a non-abelian gauge theory localized on a brane at position $y=y_0$ that interacts with the 5D scalar field through an axion-like coupling,
the corresponding brane-localized terms in the action have the form
\begin{equation}
\begin{split}
	S_{5, {\rm brane}} = \int d^4 x d y \sqrt{|g|} \frac{\delta(y - y_0)}{\sqrt{g_{55}}} & \Biggl\{ - \frac{1}{4 {\bar g}^2} g^{\mu \rho} g^{\nu \sigma} G_{\mu \nu} G_{\rho \sigma} \Biggr. \\
					& \ \ + \Biggl. \frac{\phi}{16 \pi^2 F^{3/2}} \frac{\epsilon^{\mu \nu \rho \sigma}}{\sqrt{|g_{(4)}|}} G_{\mu \nu} G_{\rho \sigma} \Biggr\} \ ,
\end{split}
\label{eq:S5scalarBrane}
\end{equation}
where $\sqrt{|g_{(4)}|} = X(y)^2$ in our notation.
The effective interaction between the gauge theory and the massless scalar mode is then
\begin{equation}
	\mathcal{L}_4	\supset \frac{ {\mathcal C}_0 \phi^{(0)} }{16 \pi^2 F^{3/2}} G_{\mu \nu} {\tilde G}^{\mu \nu}
				\equiv \frac{\phi^{(0)}}{16 \pi^2 f_0} G_{\mu \nu} {\tilde G}^{\mu \nu} \ ,
\end{equation}
where $G_{\mu \nu} {\tilde G}^{\mu \nu} \equiv \epsilon^{\mu \nu \rho \sigma} G_{\mu \nu} G_{\rho \sigma}$, as usual,
and the last expression defines an effective axion coupling $f_0$, which may be written as
\begin{equation}
	f_0 =  F^{3/2} {\mathcal C}_0^{-1} =   M_{Pl} \left( \frac{F}{M_5} \right)^{3/2} \ ,
\label{eq:f0massless}
\end{equation}
and in the last step we used the relationship between the fundamental scale of the 5D theory, $M_5$, and the 4D Planck scale $M_{Pl}$.\footnote{In a background
of the form specified in eq.(\ref{eq:metric}), this is given by $M_{Pl}^2 = M_5^3 \int_0^{\pi R} dy X(y) \sqrt{Y(y)}$.}

Eq.(\ref{eq:f0massless}) illustrates how
($i$) the effective coupling of the massless mode to the brane-localized gauge theory is independent of the position of the brane $y_0$ for any geometry --
a direct consequence of the flat profile of the zero mode -- ,
and ($ii$) a significant hierarchy between $f_0$ and $M_{Pl}$ only arises if a similar hierarchy between the 5D symmetry breaking scale $F$ and $M_5$
is introduced \emph{ad hoc} in the fundamental 5D picture.

It is illuminating to consider what happens to this theory when deconstructed.
If we latticize the extra dimension in $N$ segments, with lattice spacing $a$, such that $Na = \pi R$,
the terms in the 4D effective lagrangian corresponding to eq.(\ref{eq:S5scalar}) read\footnote{A field redefinition $\phi_j \rightarrow \phi_j (a X_j \sqrt{Y_j})^{-1/2}$ is performed to obtain canonically-normalized scalar fields.}
\begin{equation}
	\mathcal{L}_4 \supset - \frac{1}{2} \sum_{j=0}^N (\partial_\mu \phi_j)^2
			- \frac{1}{2a^2} \sum_{j=0}^{N-1} \frac{X_j}{Y_j} \left( \phi_j - \frac{X_j^{1/2} Y_j^{1/4}}{X_{j+1}^{1/2} Y_{j+1}^{1/4}} \phi_{j+1} \right)^2 \ ,
\label{eq:L4scalar}
\end{equation}
where $f_j \equiv f(aj)$ (for $f = \phi, X, Y$).
This corresponds to the effective lagrangian of eq.(\ref{eq:L4scalarDiscrete}),
with mass-squared parameters and charges
\begin{equation}
	m_j^2 = \frac{1}{a^2} \frac{X_j}{Y_j} \ , \qquad q_j = \frac{X_j^{1/2} Y_j^{1/4}}{X_{j+1}^{1/2} Y_{j+1}^{1/4}} \ ,
\label{eq:scalarDeconsParam}
\end{equation}
in agreement with what is found in~\cite{CW},
and the profile of the massless state present in the spectrum is now given by
$\phi_{(0)} = \sum_{j=0}^N c_j \phi_j$, with
\begin{equation}
	c_j = c_0 \frac{X_j^{1/2} Y_j^{1/4}}{X_0^{1/2} Y_0^{1/4}} \qquad {\rm and} \qquad
	c_0 = \left( 1 + \sum_{j=1}^N \frac{X_j \sqrt{Y_j}}{X_0 \sqrt{Y_0}} \right)^{-1/2} \ .
\label{eq:0modeScalar}
\end{equation}
In particular, the deconstruction of a real scalar field propagating in a linear dilaton background of the form $X(y) = Y(y) = e^{-4ky}$
corresponds to $m_j^2 = a^{-2}$, and $q_j = e^{3ka}$ $\forall \ j$.

However, upon deconstruction, the brane-localized terms of eq.(\ref{eq:S5scalarBrane}) read (taking into account the appropriate field redefinitions)
\begin{equation}
	\mathcal{L}_4 \supset - \frac{1}{4 {\bar g}^2} G_{\mu \nu} G^{\mu \nu}
		+ \frac{\phi_{j_0}}{16 \pi^2 F \sqrt{F a} X^{1/2}_{j_0} Y^{1/4}_{j_0}} G_{\mu \nu} {\tilde G}^{\mu \nu} \ ,
\label{eq:noMassBraneCouplingDecons}
\end{equation}
i.e.~the brane-localized interaction of the 5D theory is deconstructed into a coupling of the gauge theory to the scalar field of the $j_0$ site, where $y_0 = j_0 a$.
Written in terms of mass eigenstates, eq.(\ref{eq:noMassBraneCouplingDecons}) includes
an effective coupling between the massless scalar and the gauge sector that reads
\begin{equation}
	\mathcal{L}_4 \supset \frac{c_0}{X^{1/2}_0 Y^{1/4}_0} \frac{\phi_{(0)}}{16 \pi^2 F \sqrt{F a}} G_{\mu \nu} {\tilde G}^{\mu \nu} \ .
\label{eq:effectivef}
\end{equation}
As expected from our discussion of the continuum 5D theory, the effective axion coupling in the deconstructed theory
is independent of the position of the site $j_0$ -- in contrast with the discrete construction of section \ref{sec:discreteScalar}.
At best, conformally flat backgrounds of the linear dilaton type, for which $m_j^2 = a^{-2}$ and $q_j = q$ independent of $j$ (see eq.(\ref{eq:scalarDeconsParam})),
can \emph{accommodate} discrete clockwork, but only if the hierarchy of effective scales to be obtained in the discrete theory is put in by hand
from the 5D perspective.

\subsection{Vector case}
\label{sec:vectorNoMasses}

The action of a massless, non-interacting $U(1)$ gauge theory propagating in a non-trivial background is given by
\begin{equation}
	S_{5, {\rm bulk}} = - \int d^4 x d y \sqrt{|g|} \frac{1}{4 g^2_{5D}} g^{MR} g^{NS} F_{MN} F_{RS} \ ,
\label{eq:S5vector}
\end{equation}
where $g_{5D}$ is the 5D gauge coupling.

Working in the $A_5 = 0$ gauge, and expanding the 5D vector field as a sum over KK-modes
$A_\mu = \sum_{n = 0}^\infty \psi_n (y) A_\mu^{(n)} (x)$, the equations of motion and boundary conditions for the different modes
in the background of eq.(\ref{eq:metric}) read
\begin{equation}
	\partial_y \left( \frac{X}{\sqrt{Y}} \partial_y \psi_n \right) + m_n^2 \psi_n \sqrt{Y} = 0
\label{eq:vectorEOM}
\end{equation}
\begin{equation}
	\partial_y \psi_n = 0 \qquad {\rm at} \qquad y=0, \pi R \ .
\label{eq:vectorBC}
\end{equation}
In particular, a massless mode is present in the KK-spectrum, whose profile is a constant independent of $y$.
Without loss of generality, one may take $\psi_0 = 1$, a choice that defines
a 4D gauge coupling, $g_{4D}$, given by $g_{4D}^{-2} = g_{5D}^{-2} \int_0^{\pi R} dy \sqrt{Y(y)}$.

If we now consider a brane-localized scalar field $\varphi$ with charge $Q_\varphi$ under the $U(1)$ gauge group,
the corresponding brane-localized terms in the action read
\begin{equation}
	S_{5, {\rm brane}} = - \int d^4 x d y \sqrt{|g|} \frac{\delta(y - y_0)}{\sqrt{g_{55}}} g^{\mu \nu} (D_\mu \varphi)^\dagger D_\nu \varphi\ ,
\label{eq:S5vectorBrane}
\end{equation}
where $D_\mu \varphi = (\partial_\mu + i Q_\varphi A_\mu) \varphi$.
After the appropriate rescaling $\varphi \rightarrow \varphi / \sqrt{X (y_0)}$,
so that the scalar field features a canonically normalized kinetic term,
the terms involving $\varphi$ in the 4D effective lagrangian are given by
\begin{equation}
	\mathcal{L}_4 \supset - | (\partial_\mu + i Q_\varphi A_\mu (y_0)) \varphi |^2 = - | (\partial_\mu + i Q_\varphi A^{(0)}_\mu + ...) \varphi |^2 \ ,
\label{eq:effectiveVectorLinearDilaton}
\end{equation}
where the dots denote strictly massive vector modes.
The effective coupling between $A^{(0)}_\mu$ and the brane-localized scalar field is given by $\sim g_{4D} Q_\varphi$,
which is independent of the position of the brane along the extra dimension: Two scalar fields with the same fundamental charge under the 5D gauge theory
will couple to the massless vector mode with exactly the same strength, regardless of where they are localized --
a direct consequence of the lack of symmetry-localization of the vector zero mode.

This is consistent with what one finds upon deconstruction.
Now, from eq.(\ref{eq:S5vector}) we obtain a 4D effective lagrangian of the form
\begin{equation}
	\mathcal{L}_4 \supset - \sum_{j=0}^N \frac{1}{4 g_j^2} F^2_{j \mu \nu}
	- \frac{1}{2a^2} \sum_{j=0}^{N-1} \frac{X_j}{Y_j} \frac{1}{g_j^2} \left( A_{j \mu} - A_{j+1 \mu} \right)^2 \ ,
\label{eq:L4vector}
\end{equation}
where $g_j^{-2} = g_{5D}^{-2} a \sqrt{Y_j}$, and eq.(\ref{eq:L4vector})
corresponds to the effective lagrangian of eq.(\ref{eq:L4quiver}),
with mass-squared parameters and charges
\begin{equation}
	g_j^2 v_j^2 = \frac{1}{a^2} \frac{X_j}{Y_j} \ , \qquad q_j = 1 \ .
\label{eq:vectorClockParamDecons}
\end{equation}
In the language of section \ref{sec:symmetries}, we recognize that the linear dilaton background deconstructs into the unclockworked Theory B.

The couplings of the massless vector can be found by the substitution
$A_{j \mu} = c_0 A_{(0) \mu} + ...$, where $c_0 = 1$,
a choice that defines an effective gauge coupling for the unbroken gauge theory, $g_{(0)}$,
given by $g_{(0)}^{-2} = \sum_{j=0}^N g_j^{-2}$.
Upon deconstruction, the brane-localized terms of eq.(\ref{eq:S5vectorBrane}) read
\begin{equation}
	\mathcal{L}_4 \supset - | (\partial_\mu + i Q_\varphi A_{j_0 \mu}) \varphi |^2 = - | (\partial_\mu + i Q_\varphi A_{(0) \mu} + ...) \varphi |^2 \ ,
\end{equation}
where the dots denote strictly massive modes.
The effective coupling between the massless vector and the brane-localized scalar is $\sim g_{(0)} Q_\varphi$,
which is independent of the position where $\varphi$ is localized -- in stark contrast with the discrete theory of~\ref{sec:discreteVector}.

It becomes clear that an attempt to obtain discrete clockwork
from the deconstruction of an abelian gauge theory propagating in a non-trivial background fails, regardless of the choice of geometry.
At most, conformally flat metrics, for which $g_j^2 v_j^2 = a^{-2}$ and $q_j = 1$ when deconstructed (see eq.(\ref{eq:vectorClockParamDecons})),
can \emph{accommodate} discrete clockwork, but only if the hierarchy of effective charges between different matter fields
to be obtained in the discrete theory is put it by hand from the 5D perspective.

\subsection{Including dilaton couplings}
\label{sec:dilaton}

In theories involving a dilaton, after going from Jordan frame to Einstein frame, a $y$-dependent factor typically remains present in front of both bulk
and brane terms, and corresponds to some power of $e^S$, where $S$ is the dilaton field that gets a $y$-dependent vev.
One could wonder whether the presence of such terms alters the story told in sections \ref{sec:scalarNoMasses} and \ref{sec:vectorNoMasses},
and whether clockwork could arise from the deconstruction of theories with a dilaton.
In this section, we show that this is not the case: The presence of dilaton couplings do not qualitatively change our conclusions, so long as no additional breaking of scale invariance is introduced through the coupling of the dilaton to brane-localized states. We emphasize this requirement is a weak restriction. For instance, in the vector case, it ensures that the 5D gauge symmetry
is indeed compact. There is no symmetry localization in going from a non-compact 5D gauge symmetry to a non-compact effective 4D construction, and such models do not lead to the emergence of clockwork dynamics.

Let's consider the scalar case of section~\ref{sec:scalarNoMasses} first.
In the presence of a dilaton, eq.(\ref{eq:S5scalarBrane}) will typically include a $y$-dependent factor ${\mathcal Q} (y)$, of the form\footnote{No such factor appears,
in Einstein frame, for a bulk scalar field, and so an analogous $y$-dependent factor does not need to be included in eq.(\ref{eq:S5scalar}).}
\begin{equation}
\begin{split}
	S_{5, {\rm brane}} = \int d^4 x d y \sqrt{|g|} {\mathcal Q}(y) \frac{\delta(y - y_0)}{\sqrt{g_{55}}} & \Biggl\{ - \frac{1}{4 {\bar g}^2} g^{\mu \rho} g^{\nu \sigma} G_{\mu \nu} G_{\rho \sigma} \Biggr. \\
					& \ \ + \Biggl. \frac{\phi}{16 \pi^2 F^{3/2}} \frac{\epsilon^{\mu \nu \rho \sigma}}{\sqrt{|g_{(4)}|}} G_{\mu \nu} G_{\rho \sigma} \Biggr\} \ .
\end{split}
\label{eq:S5scalarBraneDilaton}
\end{equation}
The presence of a non-trivial function ${\mathcal Q}(y)$ alters the value of the effective gauge coupling of the brane non-abelian gauge theory,
which is now given by $g_* (y_0) = {\bar g} /  \sqrt{{\mathcal Q} (y_0)}$.
The effective interaction between the gauge theory and the massless scalar mode is modified to
\begin{equation}
	\mathcal{L}_4	\supset \left( \frac{{\bar g}}{g_* (y_0)} \right)^2 \frac{ {\mathcal C}_0 \phi^{(0)} }{16 \pi^2 F^{3/2}} G_{\mu \nu} {\tilde G}^{\mu \nu} \ ,
\end{equation}
and so the effective axion coupling is now given by
\begin{equation}
	f_0 = \left( \frac{g_*(y_0)}{{\bar g}} \right)^2 F^{3/2} {\mathcal C}_0^{-1} =  \left( \frac{g_*(y_0)}{{\bar g}} \right)^2 M_{Pl} \left( \frac{F}{M_5} \right)^{3/2} \ .
\label{eq:effectivefDilaton}
\end{equation}

From eq.(\ref{eq:effectivefDilaton}), we see that if ${\mathcal Q}(y)$ is a non-trivial function of $y$ -- a common occurrence in theories with a dilaton --
the gauge coupling of the non-abelian theory depends on $y_0$ and, in turn,
the effective axion coupling between the massless scalar and the gauge theory will depend on $y_0$ through its dependence on $g_*$.
In particular, two non-abelian gauge theories with the same fundamental gauge coupling ${\bar g}$, but localized on different branes,
will feature different effective axion couplings only because of the difference in their effective gauge couplings.
Crucially, any hierarchy in couplings involving the massless scalar field only arises as a result of the two gauge theories being physically distinct
(with different gauge couplings, and therefore different physical properties, like their confinement scales),
but not as a consequence of a symmetry-localization of the scalar zero mode.

The same effect persists when deconstructing the brane terms of eq.(\ref{eq:S5scalarBraneDilaton}).
Eq.(\ref{eq:noMassBraneCouplingDecons}) now generalizes to
\begin{equation}
	\mathcal{L}_4 \supset - \frac{1}{4 g_{* j_0}^2} G_{\mu \nu} G^{\mu \nu}
		+ \left( \frac{{\bar g}}{g_{* j_0}} \right)^2 \frac{\phi_{j_0}}{16 \pi^2 F \sqrt{F a} X^{1/2}_{j_0} Y^{1/4}_{j_0}} G_{\mu \nu} {\tilde G}^{\mu \nu} \ ,
\end{equation}
and thus the effective coupling between the massless scalar and the gauge sector reads
\begin{equation}
	\mathcal{L}_4 \supset \left( \frac{{\bar g}}{g_{* j_0}} \right)^2 \frac{c_0}{X^{1/2}_0 Y^{1/4}_0} \frac{\phi_{(0)}}{16 \pi^2 F \sqrt{F a}} G_{\mu \nu} {\tilde G}^{\mu \nu} \ .
\end{equation}
As expected from our discussion of the continuum 5D theory, the effective axion coupling in the deconstructed theory
depends on the position of the site $j_0$ only through the value of the effective gauge coupling $g_{* j_0}$.
As we move from site to site, the axion effective coupling will change as a result of the change in the properties of the non-abelian gauge theory.
This picture is in stark contrast with the clockwork mechanism described in section~\ref{sec:discreteScalar},
where the effective axion coupling changes as the gauge theory moves from site to site because of the symmetry-localization of the scalar field,
whereas the physical properties of the non-abelian gauge theory remain unchanged.

We now turn to the $U(1)$ vector case discussed in section~\ref{sec:vectorNoMasses}.
In the presence of a dilaton, eq.(\ref{eq:S5vector}) will typically include a $y$-dependent factor in front of the vector kinetic term,
of the form
\begin{equation}
	S_{5, {\rm bulk}} = - \int d^4 x d y \sqrt{|g|} \frac{{\mathcal F} (y)}{4 g^2_{5D}} g^{MR} g^{NS} F_{MN} F_{RS} \ .
\end{equation}
Although this will in general affect the equations of motion for the KK-modes, which now read
\begin{equation}
	\partial_y \left( {\mathcal F} \frac{X}{\sqrt{Y}} \partial_y \psi_n \right) + m_n^2 \psi_n {\mathcal F} \sqrt{Y} = 0 \ ,
\end{equation}
a massless mode is present in the spectrum, and its profile remains flat.
With the choice $\psi_0 = 1$, the 4D gauge coupling is now defined as $g_{4D}^{-2} = g_{5D}^{-2} \int_0^{\pi R} dy {\mathcal F} (y) \sqrt{Y(y)}$.

Similarly, eq.(\ref{eq:S5vectorBrane}) will be generalized to include a $y$-dependent factor, of the form
\begin{equation}
	S_{5, {\rm brane}} = - \int d^4 x d y \sqrt{|g|} {\mathcal H} (y) \frac{\delta(y - y_0)}{\sqrt{g_{55}}} g^{\mu \nu} (D_\mu \varphi)^\dagger D_\nu \varphi \ ,
\end{equation}
where the function ${\mathcal H} (y)$ will in general be different from ${\mathcal F} (y)$.\footnote{In models involving a dilaton, different powers of $e^S$ appear in front of
bulk and brane-localized terms when going to Einstein frame, a fact we capture here by considering two different functions ${\mathcal F} (y)$ and ${\mathcal H} (y)$.}
After the appropriate rescaling $\varphi \rightarrow \varphi / \sqrt{X (y_0) {\mathcal H} (y_0)}$,
so that the scalar field features a canonically normalized kinetic term, the terms involving $\varphi$ in the 4D effective lagrangian are just
given by eq.(\ref{eq:effectiveVectorLinearDilaton}).
The effective coupling between $A^{(0)}_\mu$ and the brane-localized scalar field is just $\sim g_{4D} Q_\varphi$ -- again independent of $y_0$.

When deconstructed, this more general case features exactly the same properties discussed in section \ref{sec:vectorNoMasses},
with the only difference that the gauge couplings on each site are now given by $g_j^{-2} = g_{5D}^{-2} a \sqrt{Y_j} {\mathcal F}_j$,
and the presence of dilaton couplings has no effect on our conclusions. The inability of dilaton couplings to reproduce meaningful clockwork is clear in the language of section \ref{sec:symmetries}: A successful modification of the linear dilaton background to generate clockwork would need to alter the physical spectrum of charged states on probe branes, rather than merely modifying gauge couplings.

\section{Towards continuum clockwork}
\label{sec:flat}

In this section, we present a 5D implementation of the clockwork mechanism that, when deconstructed,
successfully preserves the appealing features of the discrete set-up described in sections~\ref{sec:discreteScalar} and~\ref{sec:discreteVector}.
In order to emphasize how geometry plays no role, we consider the case of a flat background, and include bulk and brane mass terms for scalar and abelian vector fields.
Both from the 5D perspective, and when deconstructed, the scenario presented here features hierarchical couplings to brane-localized states
as a consequence of the symmetry-localization of the corresponding bulk fields.
In terms of the discrete clockwork parameters of sections \ref{sec:discreteScalar} and~\ref{sec:discreteVector}, the set-up we consider
appears as a small perturbation from the discrete clockwork mechanism in which all parameters are taken to be equal. We discuss the scalar case first in section~\ref{sec:scalarFlat}, albeit only at the level of a toy model; a well-defined notion of a clockworked continuum global symmetry would entail embedding the continuum global symmetry in a continuum gauge symmetry, which lies beyond the scope of the current work (see \cite{ChangSub} for work in this direction).
In section~\ref{sec:vectorFlat} we discuss the vector case in full detail, realizing a scenario in which continuum
clockwork arises when a compact 5D gauge symmetry leads to a non-compact 4D one. Section~\ref{sec:linearDilaton} clarifies the connection between our 5D construction and the  linear dilaton background implementation of~\cite{CW}.

\subsection{Continuum scalar clockwork}
\label{sec:scalarFlat}

Apart from the kinetic term of eq.(\ref{eq:S5scalar}), the 5D action of a real scalar field may also involve mass terms
\begin{equation}
	S_{5, {\rm mass}} = - \frac{1}{2} \int d^4 x d y \sqrt{|g|} 
			\phi^2 \left( M_\phi^2 + {\tilde m}_\phi \frac{\delta(y) - \delta(y - \pi R)}{\sqrt{g_{55}}} \right) \ ,
\label{eq:S5scalarMassive}
\end{equation}
where we fix the brane-localized mass terms to have equal size but opposite sign, in order to allow for a massless state
to be present in the KK-mode spectrum. Although the bulk and brane mass terms of eq.(\ref{eq:S5scalarMassive}) are certainly consistent
from an effective field theory perspective, negative brane masses might pose challenges when trying to embed this framework into a full UV completion --
an issue that we do not try to address in this work. 
In the generic warped background of eq.(\ref{eq:metric}), and after expanding the 5D scalar field $\phi$ as a sum over KK-modes as before,
the equations of motion and boundary conditions for the different modes now read
\begin{equation}
	\partial_y \left( \frac{X^2}{\sqrt{Y}} \partial_y \chi_n \right) + \chi_n X \sqrt{Y} \left(m_n^2 - X M_\phi^2 \right) = 0 \ ,
\label{eq:scalarEOMmassive}
\end{equation}
\begin{equation}
	\left(\partial_y - \frac{{\tilde m}_\phi}{2} \sqrt{Y} \right) \chi_n = 0 \qquad {\rm at} \qquad y=0, \pi R \ .
\label{eq:scalarBCmassive}
\end{equation}

As first noted in~\cite{Gherghetta:2000qt}, the presence of non-zero bulk and brane mass terms makes the zero mode's profile non-flat.
In particular, if we demand this profile to be of an exponential form $\chi_0(y) \propto e^{\beta y}$, where $\beta$ is some mass scale,
eq.(\ref{eq:scalarBCmassive}) requires $Y(y)$ is independent of $y$, and without loss of generality we may take $Y = 1$,
in which case $\beta = {\tilde m}_\phi / 2$ and thus $\chi_0(y) \propto e^{{\tilde m}_\phi y / 2}$.
Moreover, for a given $X(y)$, eq.(\ref{eq:scalarEOMmassive}) requires bulk and brane mass terms to satisfy
\begin{equation}
	{\tilde m}_\phi^2 + 4 {\tilde m}_\phi \frac{\partial_y X}{X} - 4 M_\phi^2 = 0 \ .
\label{eq:scalarBraneMass}
\end{equation}
For instance, in a flat background, where $X(y) = 1$, ${\tilde m}_\phi = \pm 2 \sqrt{M^2_\phi}$; whereas in
an RS background, where $X(y) = e^{-2 k y}$,  ${\tilde m}_\phi = 2 \left( 2k \pm \sqrt{4k^2 + M_\phi^2} \right)$,
in agreement with~\cite{Gherghetta:2000qt}.
Although choosing ${\tilde m}_\phi$ such that eq.(\ref{eq:scalarBraneMass}) is satisfied may appear like a fine-tuned choice,
we emphasize that it is a technically natural one, since only for those values of ${\tilde m}_\phi$ the lowest lying scalar mode
recovers a shift symmetry -- it is a symmetry enhanced point.
Depending on whether ${\tilde m}_\phi$ is positive or negative, the massless mode will be localized towards the $y=\pi R$ or $y=0$ branes respectively.
Here, we consider the case of a flat background ($X=Y=1$), and, without loss of generality, focus on the choice ${\tilde m}_\phi < 0$,
so that the zero mode profile is exponentially localized towards $y=0$.
(The case ${\tilde m}_\phi > 0$ is completely analogous but replaces the role of the two branes.)

In this case,
after setting ${\mathcal Q}(y) = 1$ in eq.(\ref{eq:S5scalarBrane}),
the effective axion coupling of eq.(\ref{eq:effectivef}) is given by
\begin{equation}
	f_0 = F^{3/2} \chi_0 (y_0)^{-1} = F^{3/2} \sqrt{ \frac{e^{{\tilde m}_\phi \pi R} - 1}{{\tilde m}_\phi} } e^{-{\tilde m}_\phi y_0 / 2}
		\simeq \frac{F^{3/2}}{ \sqrt{ |{\tilde m}_\phi| } } e^{| {\tilde m}_\phi | y_0 / 2} \ ,
\label{eq:effectivef0Flat}
\end{equation}
where in the last term we have focused on the case ${\tilde m}_\phi < 0$, and assumed $|{\tilde m}_\phi| \pi R = \mathcal{O}(1)$.
From eq.(\ref{eq:effectivef0Flat}), it is clear that the non-trivial profile of the zero mode translates into an effective axion coupling
that depends on the position of the brane where the gauge theory is localized.
Two gauge theories with identical properties localized on different branes will feature exponentially different effective axion couplings,
as a result of the symmetry-localization of the scalar zero mode along the extra dimension, even for natural choices of the 5D parameters, $m_\phi \pi R = \mathcal{O} (1)$.

We now consider the deconstruction of this theory, and compare it to the discrete clockwork of section~\ref{sec:discreteScalar}.
The 4D effective lagrangian of the scalar sector reads
\begin{equation}
	\mathcal{L}_4 = - \frac{1}{2} \sum_{j=0}^N (\partial_\mu \phi_j)^2 - \frac{1}{2} \sum_{i, j=0}^N \mathbb{M}_{\phi, ij}^2 \phi_i \phi_j \ ,
\end{equation}
with a mass-squared matrix given by
\begin{equation}
\begin{split}
	\mathbb{M}^2_{\phi, i j} =
			&\delta_{i j} \left( M_\phi^2 + \frac{2}{a^2} \right)
			- \frac{1}{a^2} (\delta_{i j+1} +  \delta_{i j-1}) \\
			- &\delta_{iN} \delta_{jN} \left( \frac{{\tilde m}_\phi}{a} + \frac{1}{a^2} \right)
			+ \delta_{i0} \delta_{j0} \left( \frac{{\tilde m}_\phi}{a} - \frac{1}{a^2} \right) \ .
\end{split}
\label{eq:flatMassScalar}
\end{equation}
As in the continuum case, for a given bulk mass term $M_\phi^2$ there are two values of ${\tilde m}_\phi$ that allow
for a massless mode to be present in the latticized spectrum, which are of equal size but opposite sign,\footnote{One can check that ${\tilde m}_\phi = \pm \left( 2 \sqrt{M_\phi^2} + \mathcal{O} (1/N) \right) $, as expected.} and,
as before, we focus on the case ${\tilde m}_\phi < 0$.

Moreover, upon deconstruction the brane-localized coupling between the 4D gauge theory and the 5D scalar field now reads
\begin{equation}
	\mathcal{L}_4 \supset \frac{\phi_{j_0}}{16 \pi^2 F \sqrt{F a}} G_{\mu \nu} {\tilde G}^{\mu \nu} = \frac{c_{j_0} \phi_{(0)}}{16 \pi^2 F \sqrt{F a}} G_{\mu \nu} {\tilde G}^{\mu \nu} + ... \ ,
\label{eq:flatBraneCouplingDecons}
\end{equation}
where the dots correspond to strictly massive modes, and
the effective axion coupling scale is now given by $f_0 = F \sqrt{Fa} \ c_{j_0}^{-1}$.
Unlike the scenarios considered in section~\ref{sec:scalarNoMasses},
the deconstructed effective coupling now depends on $j_0$ as a result of the uneven distribution of the massless scalar
along the different lattice sites, mirroring the situation found from the 5D perspective.

\begin{figure}
\subfigure{\includegraphics[scale=0.5]{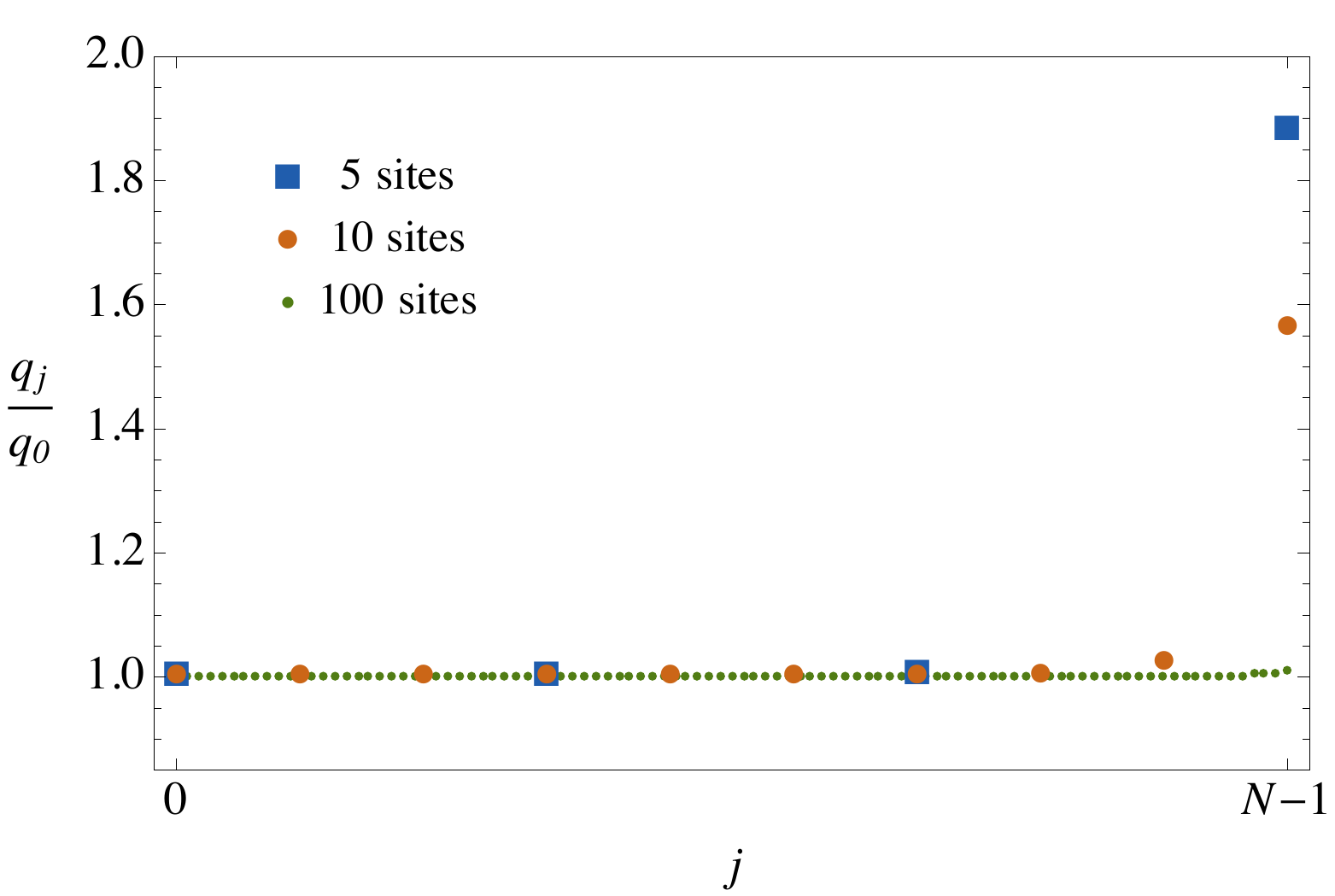}}
\hfill
\subfigure{\includegraphics[scale=0.5]{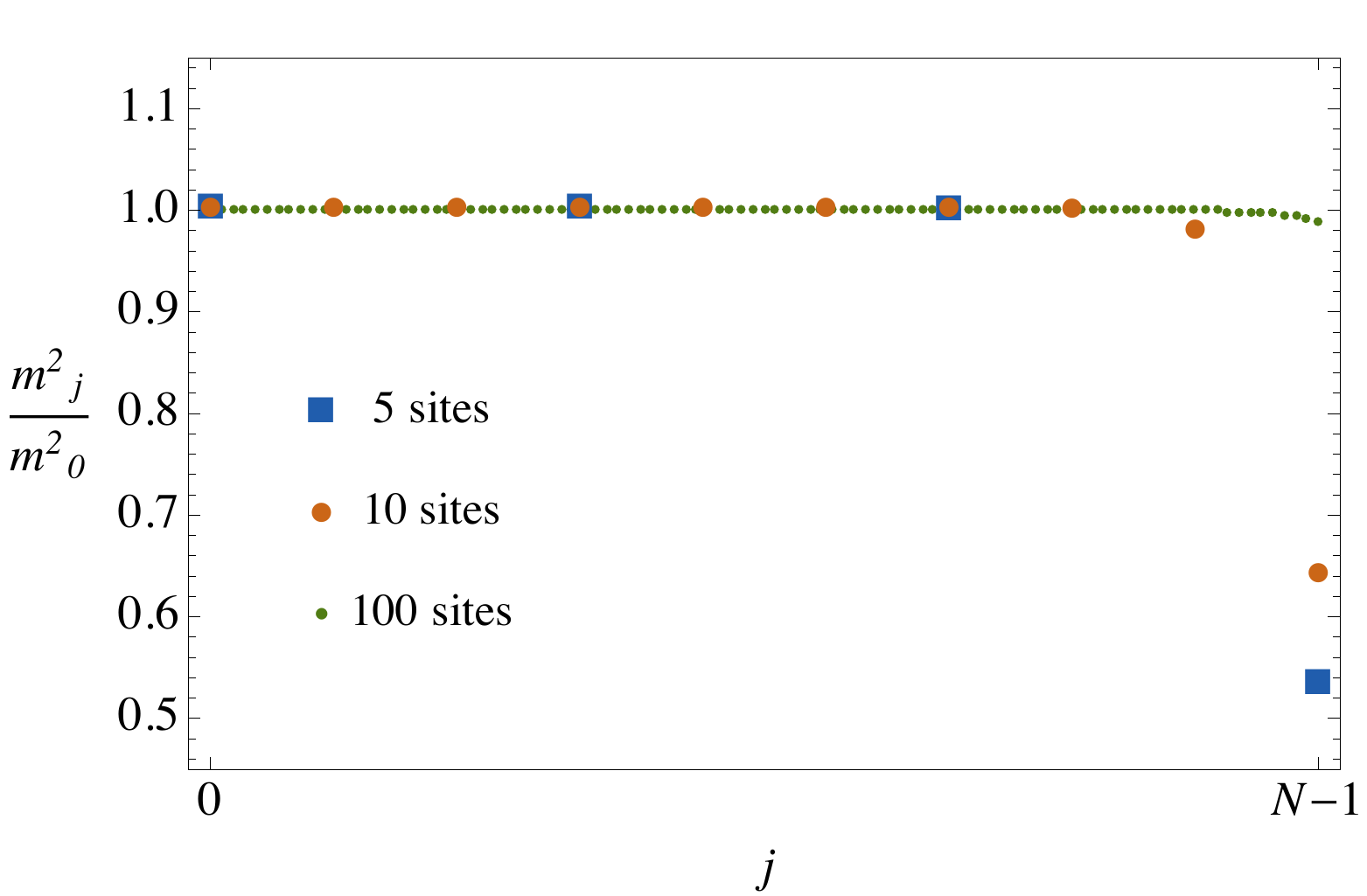}}
\caption{\textbf{Left:} Values of the charges $q_j$ (normalized to $q_0$, for $j=0,...,N-1$) that correspond
to a discrete clockwork mechanism arising from the deconstruction of a massive 5D scalar field in a flat background, as described in section~\ref{sec:scalarFlat}.
For illustration, we choose $\sqrt{M_\phi^2} \pi R =15$, and focus on the case of $N=5, 10$, and 100 lattice sites.
We make the first and last point coincident, so that the hierarchy between the first and last charge parameters,
and how it changes as we increase the number of sites, be compared between all three cases.
\textbf{Right:} Same as in the left figure but for the mass-squared clockwork parameters $m_j^2$.}
\label{fig:flatCWparameters}
\end{figure}

\begin{figure}
\centering
\includegraphics[scale=0.5]{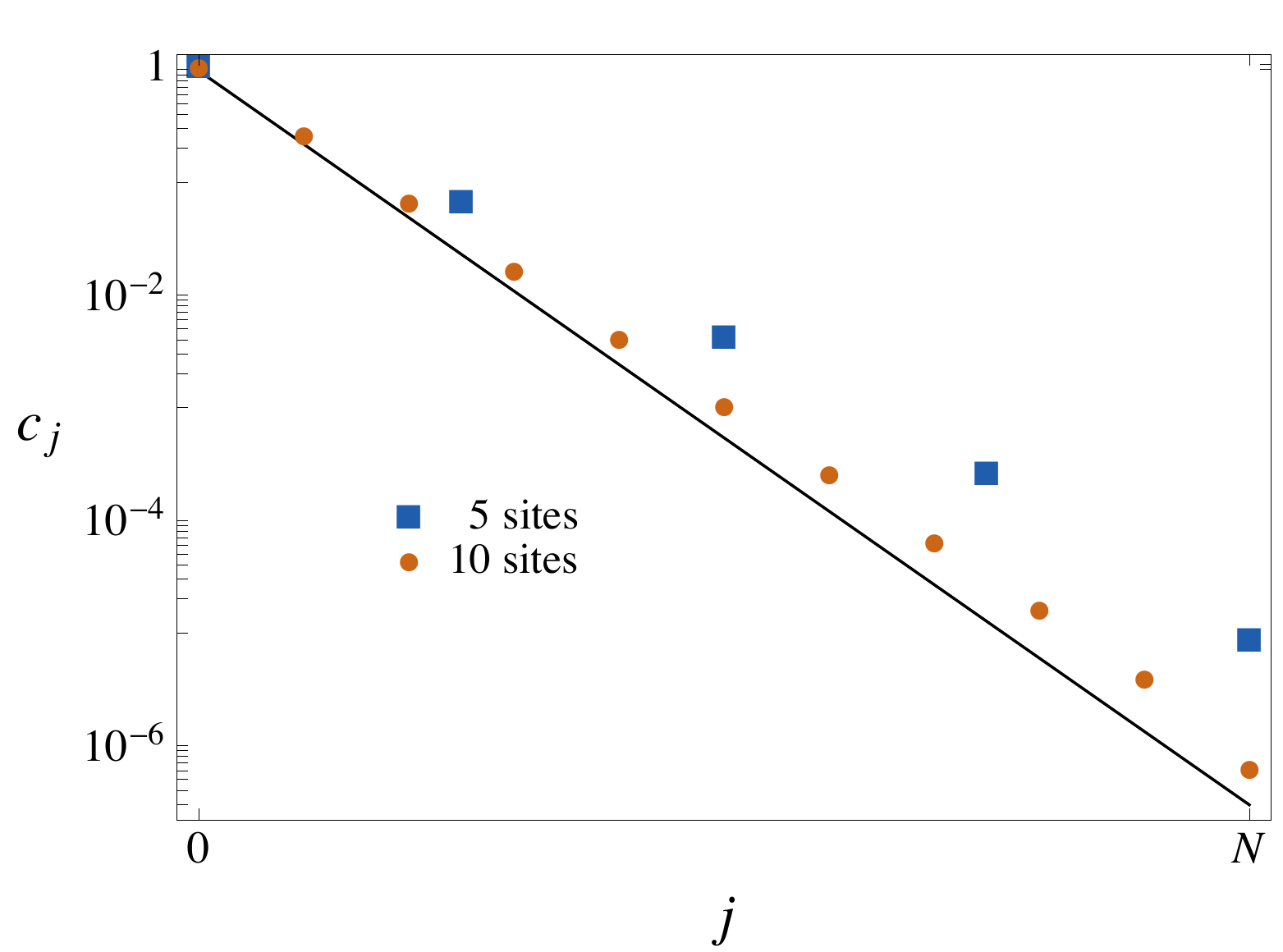}
\caption{Profile of the massless mode obtained from the deconstruction of a massive 5D scalar field in a flat background,
as described in section~\ref{sec:scalarFlat},
for the case of $N=5$ and 10 lattice sites, and for $\sqrt{M_\phi^2} \pi R = 15$ for illustration.
The black line corresponds to the case of an exact exponential profile $\propto e^{ - \sqrt{M_\phi^2} y}$.
}
\label{fig:flatCWprofile}
\end{figure}

In particular, it is illuminating to match this deconstructed scenario into the discrete clockwork set-up of section~\ref{sec:discreteScalar},
by finding the corresponding clockwork parameters
$q_i$ and $m_i^2$ ($i=0,...,N-1$), since one may worry that this may now look like an unnaturally hierarchical set of choices,
and that the `naturalness' we recover in the 5D picture by introducing mass terms and considering a flat background, may be lost in the deconstruction.
Instead, we find that this is not the case: When deconstructed, the scenario we consider has approximately equal $q_j$ and $m_j^2$ parameters.
To illustrate this fact, in figure~\ref{fig:flatCWparameters} we show the values of $q_i$ and $m_i^2$ (normalized to the values on the first site)
for $\sqrt{M_\phi^2} \pi R = 15$ (just for illustration).
From figure~\ref{fig:flatCWparameters} one can appreciate that the effective charges and mass-squared parameters are all of similar size,
no large hierarchies between them are present, and all of them tend to the same value as one approaches the large $N$ limit.
As a result, the profile of the massless mode also very closely resembles an exponential,
as we illustrate in figure~\ref{fig:flatCWprofile}.

\subsection{Continuum vector clockwork}
\label{sec:vectorFlat}

As in the scalar case discussed in the previous section, we may in general include both bulk and brane mass terms for a 5D
abelian gauge field,\footnote{The unconventional $g_{5D}^{-2}$ factor in front of eq.(\ref{eq:S5vectorMassive}) is in keeping with our notation in previous sections,
and in particular with eq.(\ref{eq:S5vector}).}
\begin{equation}
	S_{5, {\rm mass}} = - \frac{1}{2 g_{5D}^2} \int d^4 x d y \sqrt{|g|} g^{MN} A_M A_N \left( M_A^2 + {\tilde m}_A \frac{\delta (y) - \delta (y-\pi R)}{\sqrt{g_{55}}} \right) \ ,
\label{eq:S5vectorMassive}
\end{equation}
and we note that these may be generated through spontaneous symmetry breaking
(due to the non-zero vev of a 5D scalar field featuring both bulk and brane-localized kinetic terms, as pointed out in~\cite{Kogan:2001wp}),
and thus do not necessarily require an explicit breaking of the fundamental 5D gauge symmetry. As in section \ref{sec:scalarFlat}, the presence
of negative brane-localized mass terms is consistent within an effective description, although such terms may be non-trivial to realize in the context of a full UV completion and potentially pose an obstruction to genuine continuum clockwork.\footnote{In particular, the UV completion of these mass terms in a compact higher-dimensional gauge theory requires the introduction of a bulk Higgs field with wrong-sign brane kinetic terms. Apart from the challenges posed by the brane kinetic terms, avoiding significant back-reaction on the geometry from the bulk Higgs vev implies $M_A \ll M_5$. This still permits a meaningful exponential profile provided the extra dimension is stabilized with $R M_5 \gg 1$ but precludes arbitrarily strong symmetry localization. We thank M.~McCullough for discussion of this point.}

As was first noted in~\cite{Kogan:2001wp,Ghoroku:2001zu,Batell:2005wa}, the presence of non-zero bulk and brane mass terms makes the
profile of the vector zero mode non-flat.
For an exponential profile $\psi_0(y)$,
the appropriate boundary conditions require again $Y = 1$, in which case $\psi_0(y) \propto e^{{\tilde m}_A y / 2}$.
Without loss of generality, one may take $\psi_0(y) = e^{{\tilde m}_A y / 2}$, a choice that defines a 4D gauge coupling $g_{4D}$
given by $g_{4D}^{-2} = g_{5D}^{-2} (e^{{\tilde m}_A \pi R} - 1)/{\tilde m}_A$.
Moreover, for a given $X(y)$, the equation of motion for the zero mode demands bulk and brane mass terms to satisfy
\begin{equation}
	{\tilde m}_A^2 + 2 {\tilde m}_A \frac{\partial_y X}{X} - 4 M_A^2 = 0 \ .
\label{eq:vectorBraneMass}
\end{equation}
For instance, in a flat background, where $X(y) = 1$, ${\tilde m}_A = \pm 2 \sqrt{M^2_A}$;
whereas in an RS background, where $X(y) = e^{-2 k y}$,  ${\tilde m}_A = 2 \left( k \pm \sqrt{k^2 + M_A^2} \right)$
(in agreement with~\cite{Batell:2005wa}).
As before, the values of ${\tilde m}_A$ that satisfy eq.(\ref{eq:vectorBraneMass}) constitute a technically natural choice of parameters,
since only for those values the theory recovers 4D gauge invariance of the zero mode -- again, a symmetry protected choice.

In the case of a flat background ($X=Y=1$),
the effective interaction term defined through eq.(\ref{eq:effectiveVectorLinearDilaton}) is now given by
\begin{equation}
 \mathcal{L}_4 \supset - | (\partial_\mu + i Q_\varphi A_\mu (y_0)) \varphi |^2 = - | (\partial_\mu + i Q_\varphi e^{{\tilde m}_A y_0 / 2} A_{(0) \mu} + ...) \varphi |^2 \ .
\label{eq:4dVectorBraneCoupling}
\end{equation}
The effective coupling between the scalar field and the massless vector is $\sim g_{4D} Q_\varphi e^{{\tilde m}_A y_0 / 2}$.
An exponential hierarchy of effective charges may now be generated by localizing matter on opposite branes, as a result of the physical
localization of the vector zero mode.

When deconstructed, the general features of the discrete version are very similar to those of the scalar case
described in section~\ref{sec:scalarFlat}.
For finite $N$, the discrete clockwork parameters all have similar size, and asymptote to a common value in the continuum limit,
whereas the distribution of the massless mode along the different sites approaches again an exponential profile.

One may try to implement an analogous mechanism for a non-abelian gauge theory.
This possibility was considered in \cite{Batell:2006dp}, where both bulk and brane mass terms (of the right size)
are included for a non-abelian gauge theory propagating in a slice of AdS,
and the authors of \cite{Batell:2006dp} find that an exponentially-localized zero mode is present in the KK-spectrum.
As a result of the zero mode's non-trivial profile, its cubic and quartic couplings are found to differ,
and brane-localized kinetic terms need to be included to render them equal.
This ensures that those terms in the effective lagrangian involving \emph{only} the massless vector mode exhibit 4D gauge
invariance.
However, gauge invariance of the zero mode also requires interaction terms involving the zero mode and massive KK-modes
be gauge invariant independently -- a requirement that is not fulfilled in \cite{Batell:2006dp}.
As a result, although the lowest-lying vector mode appears massless at tree-level, its mass remains unprotected under  quantum corrections.

\subsection{Relation to linear dilaton theories}
\label{sec:linearDilaton}

As we have seen in sections \ref{sec:scalarNoMasses} and \ref{sec:scalarFlat}, the scalar clockwork parameters
corresponding to the deconstruction of a massive scalar field propagating
in a flat background are rather similar to those that arise in the deconstruction of a massless field in a linear dilaton geometry.
In terms of the discrete clockwork mechanism of section~\ref{sec:discreteScalar}, the latter seem to correspond to
identical clockwork parameters across sites, whereas the former appears just as a small perturbation thereof.

The reason for this similarity is a deeper relation between the two theories at the 5D level.
The KK-mode spectrum of a massless 5D scalar theory in a background given by functions $X(y)$ and $Y(y)$
is identical to that of a massive theory with an exponentially localized zero mode, $\chi_0(y) \propto e^{{\tilde m}_\phi y /2}$,
in a background given by functions ${\tilde X}(y)$ and ${\tilde Y}(y) = 1$, provided
\begin{equation}
	Y(y) = e^{2 {\tilde m}_\phi y / 3} \ , \qquad {\rm and} \qquad X(y) = {\tilde X} (y) Y(y) = {\tilde X} (y) e^{2 {\tilde m}_\phi y / 3} \ .
\end{equation}
(This can be checked from eq.(\ref{eq:scalarEOMmassive}) and eq.(\ref{eq:scalarBCmassive})
by performing a field redefinition $\chi_n \rightarrow e^{{\tilde m}_\phi y / 2} \chi_n $, and taking into account eq.(\ref{eq:scalarBraneMass}).)
Whereas the two theories are identical as far as the scalar sector is concerned (the spectrum of KK-mode masses is the same), the profiles of the different modes
in the massive theory correspond to those of the massless theory after a rescaling by a factor of $e^{{\tilde m}_\phi y / 2}$.
This feature crucially distinguishes the two theories when the 5D scalar couples to brane-localized states: In the flat case,
the massless mode is symmetry-localized along the extra dimension, whereas this is not the case in linear dilaton geometries.
Only in the flat case, couplings between the scalar zero mode and brane-localized states depend exponentially on the position of the branes
as a result of the zero mode's non-trivial profile.

In particular, for a massless scalar field in a linear dilaton geometry $X(y) = Y(y) = e^{-4 k y}$, the spectrum of KK-modes is identical
to that of a massive theory with ${\tilde m}_\phi = - 6 k$ (therefore $M_\phi^2 = (3k)^2$),
and $\tilde Y = \tilde X = 1$ -- i.e.~a massive scalar theory in a flat background, of the kind considered in section~\ref{sec:scalarFlat}.
In the continuum limit, the mass spectrum of KK-modes is given by
\begin{equation}
	\frac{m_n^2}{(1/R)^2} = (3 k R)^2 + n^2 = M_\phi^2 R^2 + n^2 \qquad {\rm for} \ n \geq 1 \ .
\end{equation}
Upon deconstruction, the spectrum of massive states, i.e.~the `clockwork gears', differ between the two theories for a given number of sites $N$.
In particular, in the linear dilaton background with vanishing bulk and brane masses, the mass of the $n$-th clockwork gear is given by~\cite{CW}
\begin{equation}
	\left. m_n^2 \right|_{\rm L.D} = m^2 \left( q^2 + 1 - 2q \cos\frac{\pi n}{N+1} \right) \ ,
\end{equation}
with $q = e^{3ka}$ and $m^2 = a^{-2}$.
Taking the large $N$ limit, while keeping $N a = \pi R$ constant,
\begin{equation}
	\left. \frac{m_n^2}{(1/R)^2} \right|_{\rm L.D} = (3 k R)^2 + n^2 + \frac{1}{N} \left( \frac{27}{\pi^2} (k \pi R)^3 + n^2 ( 3 k \pi R - 2) \right) + \mathcal{O} \left( \frac{1}{N^2} \right) \ ,
\label{eq:gearsLinearDilaton}
\end{equation}
and so the mass of the $n$-th clockwork gear approaches the mass of the $n$-th KK-mode linearly in $1/N$.
Similarly, when deconstructing the theory of section~\ref{sec:scalarFlat}, the mass of the $n$-th clockwork gear approaches the mass of the $n$-th KK-mode linearly in $1/N$,
although the size of the $\sim 1/N$ corrections in the flat deconstruction is much smaller than in the linear dilaton deconstruction.\footnote{For instance,
for $k \pi R = 5$, the mass of the first clockwork gear in the linear dilaton background only comes to within $10 \%$ of the first KK-mode
for $N \approx 80$. On the other hand, the deconstruction of the flat theory already features the first clockwork gear with a mass less than $2\%$
different from that of the first KK-mode in the most minimal case of three lattice sites ($N=2$).}
In any case, both deconstructions reproduce the same mass matrix for the scalar sector up to $1/N$ corrections.

Crucially, however, the symmetry-localization of the zero mode in the theory of section \ref{sec:scalarFlat}, and the absence of it in linear dilaton theories,
leads to similarly different behaviour upon deconstruction: Whereas the deconstruction of the theory in flat space with bulk and brane masses leads to
a meaningful clockwork mechanism (i.e.~it corresponds to Theory A, in the language of section \ref{sec:symmetries}),
deconstructing the linear dilaton theory merely leads to a discrete theory with approximately the same spectrum of massive modes, but it does \emph{not} exhibit clockwork dynamics (i.e.~it corresponds to Theory B).

\begin{figure}
\centering
\includegraphics[scale=0.65]{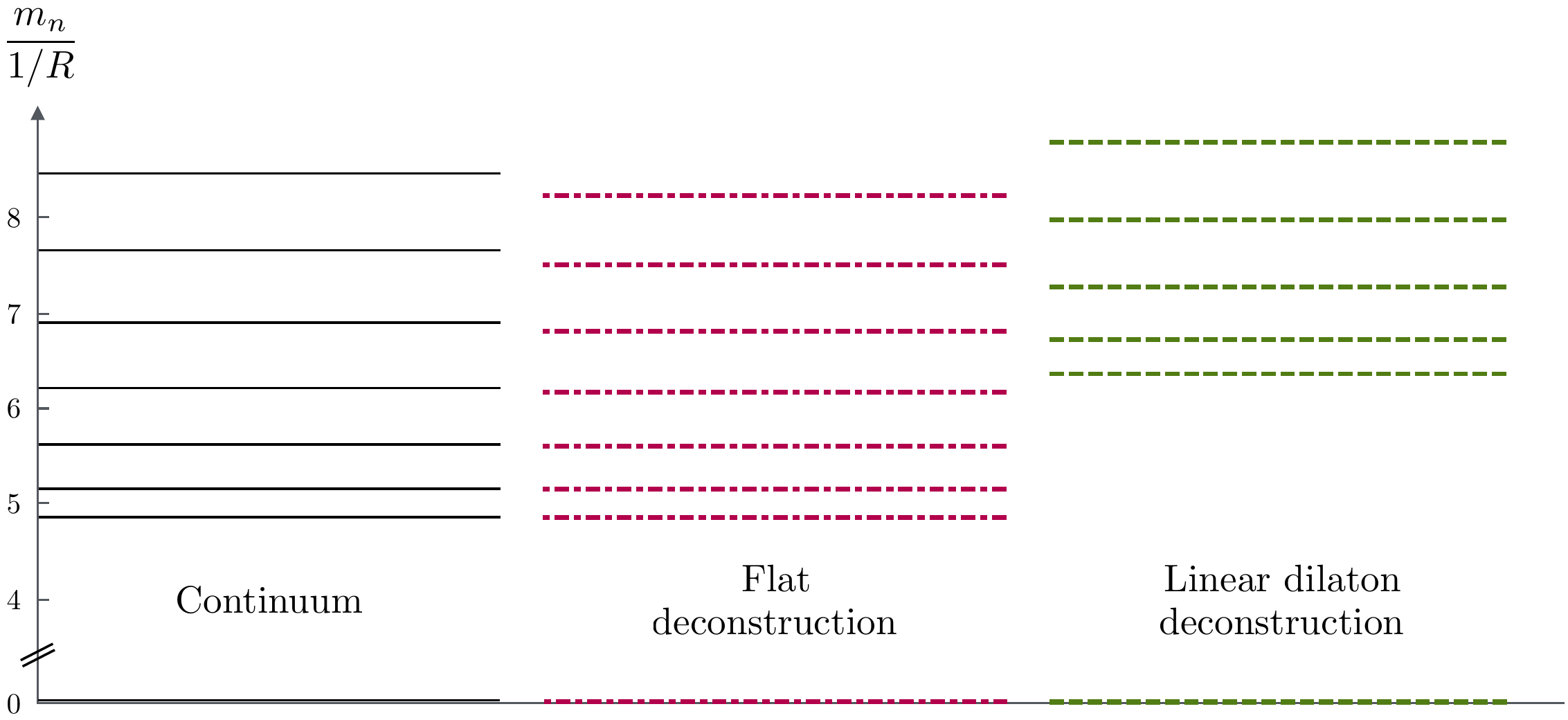}
\caption{{\bf Left:} Spectrum of massive KK-modes for a massless scalar field in a linear dilaton geometry given by $X(y) = Y(y) = e^{-4 k y}$,
which is identical to that of a scalar field in a flat background with bulk mass $M_\phi^2 = 3 k$, and boundary mass parameter ${\tilde m}_\phi = -6k$.
{\bf Center:} Spectrum of clockwork gears arising from the deconstruction of a massive scalar field in a flat background.
{\bf Right:} Spectrum of clockwork gears arising from the deconstruction of a massless scalar field in a linear dilaton geometry.
For illustration, we take  $k \pi R = 5$ in all three cases, and the center and right figures correspond to deconstructions featuring $N=30$ sites.
}
\label{fig:gears}
\end{figure}

The situation for the vector case is completely analogous to the scalar case described above.
The KK-mode spectrum of a massless 5D $U(1)$ gauge theory in a background given by $X(y)$ and $Y(y)$
is identical to that of a massive theory with an exponentially localized zero mode, $\psi_0(y) \propto e^{{\tilde m}_A y / 2}$, in a background given by
$\tilde X(y)$ and $\tilde Y(y) = 1$, provided
\begin{equation}
	Y(y) = e^{2 {\tilde m}_A y} \ , \qquad {\rm and} \qquad X (y) = \tilde X(y) Y(y) = \tilde X(y) e^{2 {\tilde m}_A y} \ .
\end{equation}
As before, although the spectrum of masses is the same in the two theories, the mode profiles differ by an overall factor of $e^{{\tilde m}_A y / 2}$,
and therefore the two theories exhibit crucially different behavior in their couplings to brane-localized states.
In particular, the spectrum of KK-modes of
a massless $U(1)$ gauge field in a linear dilaton background is identical to that of a massive theory with ${\tilde m}_A = - 2 k$, and
$\tilde Y = \tilde X = 1$.
Upon deconstruction, the spectrum of massive vector modes will be the same up to $1/N$ corrections, but, just as in the scalar case discussed above,
only one of the theories exhibits meaningful clockwork dynamics -- that of a massive 5D vector with bulk and brane masses in a flat background.

\section{Conclusions}
\label{sec:conclusions}

The elusiveness of physics beyond the Standard Model strongly motivates the search for theories in which large hierarchies of effective interactions arise from natural fundamental parameters. The clockwork mechanism beautifully realizes this goal, generating exponentially-suppressed couplings to a symmetry-protected state without significant hierarchies in the ultraviolet theory. Such a phenomenon invites both exploration of its full scope and application to extensions of the Standard Model. 

In this paper we have systematically investigated the scope of clockwork phenomena in four dimensions, as well as possible continuum counterparts in five dimensions. We have demonstrated that clockwork is an intrinsically abelian phenomenon, suitable for generating exponentially suppressed couplings to goldstone bosons of spontaneously broken abelian global symmetries, or to gauge bosons of abelian local symmetries. It is manifestly impossible to realize a clockwork mechanism for non-abelian symmetries protecting a light state, precluding the application of clockwork to Yang-Mills theories, non-linear sigma models, or gravity (thereby frustrating any attempts to solve the hierarchy problem by clockworking gravity). We illustrate this with both explicit four-dimensional constructions and more general group-theoretic arguments. 

We have also explored the extent to which viable clockwork models in four dimensions have continuum counterparts in five dimensions. We study a general class of five-dimensional theories with a compact fifth dimension, whose metrics preserve four-dimensional Lorentz invariance with warp factors that are a function of the fifth coordinate. Members of the class include flat, Randall-Sundrum, and linear dilaton models. The zero modes of all massless bosonic bulk fields on these metrics are flat in the sense that they couple equally to states localized on codimension-one surfaces anywhere in the fifth dimension. These five-dimensional theories are therefore not continuum counterparts of four-dimensional clockwork. Moreover, their deconstructions cannot be identified with four-dimensional clockwork, as zero modes in the deconstructions couple universally to states localized at specific sites, in contrast with clockwork. In addition, any nontrivial warp factor in the metric of a higher-dimensional theory corresponds to a hierarchy of couplings and scales intrinsic to each site in its deconstruction, again in contrast with clockwork.\footnote{For abelian bulk fields the hierarchies of scales and couplings in the deconstruction can be absorbed into genuine clockwork-like charges, but the zero mode in these deconstructions still lacks the position-dependent couplings of  clockwork unless position-dependent charges are put in by hand at the outset.} 

Among other things, this implies that linear dilaton models (and more generally a broad class of five-dimensional theories whose metrics give nominal hierarchies) are not the continuum counterparts of clockwork. Linear dilaton theories may still be of interest in addressing the hierarchy problem in their own right, but they so in a way that is unrelated to clockwork. In particular, the deconstruction of gravity in linear dilaton backgrounds necessarily involves the same sort of site-by-site scale hierarchies found in the deconstruction of Randall-Sundrum models, rather than the parametrically similar scales found in clockwork. 

This leaves the question of what five-dimensional theories, if any, {\it are} the continuum counterparts of abelian clockwork models. Although physically meaningful coupling hierarchies for the zero modes of bulk bosons cannot be generated by metric factors, they {\it can} be generated by non-trivial zero mode profiles unrelated to the metric. We have found that candidate continuum counterparts of abelian clockwork involve scalars or vectors with bulk and brane masses tuned to preserve a massless zero mode. This imparts a physically meaningful profile to the zero mode that generates the desired exponential and position-dependent hierarchy in couplings to localized states. Deconstructions of these continuum theories do exhibit clockwork phenomena, and their masses and couplings agree with those of uniform clockwork up to corrections that fall off with the number of sites. These five-dimensional theories may be a fruitful setting for additional clockwork model-building.

\section*{Note added}
Shortly after our work appeared, a comment on its content was made by Giudice and McCullough \cite{Giudice:2017suc}. As \cite{Giudice:2017suc} does not dispute the main technical results of this work, we present some brief remarks regarding the results' interpretation.

\begin{itemize}
\item In \cite{Giudice:2017suc}, the authors advocate a restriction of clockwork to the effective theory, treating the couplings of the zero mode as ultraviolet properties beyond the scope of the mechanism. We note however that the generation of ``exponentially suppressed interactions in theories which contain no small parameters at the fundamental level'' (a characterization of clockwork supported by \cite{CW,Giudice:2017suc}) is an inherently ultraviolet-sensitive question. Restricted to an effective theory, the naturalness constraints on the sizes of charges, couplings, or field excursions that motivated the original clockwork models \cite{Choi:2015fiu,Kaplan:2015fuy, Saraswat:2016eaz} simply do not exist, and clockwork is an unnecessary mechanism. Moreover, when restricted to an effective theory there is no meaningful notion of ``exponentially suppressed interactions in theories which contain no small parameters at the fundamental level''. For example, in the abelian case the fact that Theory A and Theory B are indistinguishable without reference to the charge spectrum or other ultraviolet properties of the theory implies that there is no invariant notion of exponential hierarchies without small parameters in the effective theory alone. This highlights the fact that, as we have argued, ultraviolet properties of the theory are necessary to a meaningful definition of clockwork.

\item A primary concern of \cite{Giudice:2017suc} is whether the continuum scalar toy model presented in section \ref{sec:scalarFlat} is simply a field redefinition of the scalar in a linear dilaton background presented in \cite{CW}. Absent a well-defined notion of a compact symmetry for the scalar, this is of course true, but in explicit models where there is a clear notion of a compact symmetry -- as would be the case for scalars arising as the fifth component of a bulk vector field -- invariant distinctions exist.  For a concrete construction along these lines, see \cite{ChangSub}. The toy model in section \ref{sec:scalarFlat} is intended to exemplify the properties that such a continuum clockwork theory should possess, namely a mechanism for altering the profile of the zero mode relative to a natural basis of gauge or global symmetry charges. 

\item We emphasize that the notion of clockworked symmetries for fields in linear dilaton backgrounds presented in \cite{Giudice:2017suc} is equivalent to our observation that the unbroken symmetry group protecting the zero mode is always the diagonal one, and the input couplings vary exponentially.
This can be obscured by changing the normalizations of generators or transformation parameters, but does not alter the essential symmetry structure at the level of the unbroken subgroup. 
\end{itemize} 

In light of the above, we persist in our position that meaningful clockwork along the lines of the original works \cite{Choi:2015fiu,Kaplan:2015fuy, Saraswat:2016eaz} is an inherently ultraviolet-sensitive mechanism, and that its key property is the preservation of an asymmetrically-distributed unbroken symmetry subgroup. However, being physicists and not linguists, we do not wish to constrain those who would prefer to adopt a different definition. For those who prefer a big-tent notion of clockwork that encompasses all theories with the same mass matrix and incommensurable symmetry properties ({\it a la} \cite{CW, Giudice:2017suc}), we might advocate referring to theories with an asymmetrically-distributed unbroken symmetry subgroup (such as \cite{Choi:2015fiu,Kaplan:2015fuy, Saraswat:2016eaz}) as ``tourbillon clockwork,'' since they are protected against ultraviolet restrictions on the natural size of parameters, such as those imposed by gravity. Theories with a symmetrically-distributed unbroken symmetry subgroup (such as those arising in linear dilaton backgrounds \cite{CW})  might correspondingly be distinguished as ``non-tourbillon clockwork.'' We hope that nature affords us clockwork of either kind.

\acknowledgments
We thank Asimina Arvanitaki, Masha Baryakhtar, Patrick Draper, Nemanja Kaloper, Alex Pomarol, Riccardo Rattazzi, Diego Redigolo, and Tomer Volansky for useful discussions.
The authors are particularly grateful to James Bonifacio, Tony Gherghetta,
Ben Gripaios, Kiel Howe, and John March-Russell for detailed comments on the manuscript.
We would also like to thank Gian Giudice and Matthew McCullough for constructive dialogue about clockwork.
IGG thanks the Kavli Institute for Theoretical Physics for the award of a graduate visiting fellowship.
The work of NC and DS is supported in part by the US Department of Energy under the grant DE-SC0014129.
IGG is financially supported by the British STFC, and, while at KITP, by the Gordon and Betty Moore Foundation under grant no.~4310.
This research was supported in part by the National Science Foundation under grant no.~NSF PHY11-25915.

\bibliography{clockworkRefs}

\appendix

\section{More group theory of clockwork\label{app:group}}

We wish to find the subgroups $G < G^N$, i.e. we wish to identify a subset of elements of $G^N$ which, under $G^N$ composition rules, form a subgroup isomorphic to $G$. We may generically write the subgroup's elements as
\begin{equation}
\lbrace (\varphi_1(g),\varphi_2(g),\ldots,\varphi_N(g)) | g \in G \rbrace ,
\end{equation}
where $\varphi_i: G \to G_i$ are maps from the subgroup $G$ to the $i$th group in the $G^N$ product, and collectively the $\varphi_i$ identify a $G^N$ tuple corresponding to each element of $G$. The $\varphi_i$ respect group composition $\varphi_i(g) \varphi_i(g^\prime) = \varphi_i(g g^\prime)$; each $\varphi_i$ is therefore a group homomorphism. The kernel of each homomorphism $\varphi_i$ must be a normal subgroup of $G$, and the image of $\varphi_i$ in $G_i$ must be the corresponding quotient group $G/\ker \varphi_i$. Moreover, the quotient group must also be a subgroup of $G_i$.

For any group $G$, there are thus at least two options for the map $\varphi_i$. One, $\ker \varphi_i = G$ and all elements of $G$ are mapped to the identity of $G_i$. Two, $\ker \varphi_i = \{e\}$, the trivial group, and $\varphi_i$ is an isomorphism from $G$ to $G_i$. For there to be a third option, we require both that $G$ has a non-trivial normal subgroup $K \equiv \ker \varphi_i$, and that the quotient group $H \equiv G/K$ is also a subgroup of $G$. This occurs, for instance, if we can write $G$ as a non-trivial semidirect product $G = K \rtimes H$.\footnote{By categorizing the behaviors of $\varphi_i$ in an analogous way, it is possible to enumerate the subgroups $S < G^N$, given $S<G$. There are two senses in which the possibilities for $S$ may be greater. One, there can be $S<G^N$ which do not result from the restriction of any $G<G^N$. This occurs either if there exists a $K \triangleleft S$ which is not normal in $G$, or if there exists a $K$ where $S/K < G_i$ but $G/K$ is not a subgroup of $G_i$. Two, there can be $S<G^N$ which do not satisfy $S<S^N$. This occurs if there exists an $S/K < G_i$ which is not a subgroup of $S_i$.}

Finally note that, so that all elements of $G$ are mapped to distinct elements of $G^N$, we require that the image of any two distinct group elements is distinct under at least one of the $\varphi_i$ (i.e. $\forall g,g^\prime \in G, g \neq g^\prime \implies \exists i, \varphi_i(g) \neq \varphi_i(g^\prime)$).

This result is effectively an iterative application of Goursat's lemma \cite{Goursat1889,2011arXiv1109.0024B}, which enumerates in full generality the subgroups of any direct product group. The result is also rather restrictive: for instance, the classical Lie groups only have discrete normal subgroups, and none of the corresponding quotient groups are subgroups. For a classical Lie group, each $\varphi_i$ is therefore either a group isomorphism or a map to the trivial group, and at most $N-1$ of them may fall in the latter category. Colloquially, we can break some --- but not all --- of the $G^N$ entirely, and we must then break `to the diagonal' of the remaining parts of $G^N$.

Thus, consider an $N$-site quiver, where each site has a symmetry group $G$ with no third option for the $\varphi_i$. If we arrange symmetry breaking interactions such that $G^N \to G$, the remaining $G$ will have equal action at any two sites of the lattice (up to isomorphisms, and assuming both sites transform non-trivially). Note that this result does not fix the normalization of fields transforming under representations of $G^N$. For instance, if $\phi_1$ and $\phi_2$ are complex scalars transforming in the same unirrep of the left-hand and right-hand parts respectively of $G \times G$, the lagrangian
\begin{equation}
\mathcal{L} = -\abs{\partial \phi_1}^2 -\abs{\partial \phi_2}^2 - \epsilon \abs{\phi_1 - q \phi_2}^2,
\end{equation}
explicitly breaks the $G \times G$ symmetry of the kinetic terms to the diagonal, whereas the massless mode $\propto q \phi_1 + \phi_2$ is weighted asymmetrically over the original fields. Importantly, however, this mode is not forced to be massless by the remaining $G$ symmetry, as a mass term formed by the unirrep scalar and its conjugate is always invariant thereunder.

There are important classes of models where the normalization of the fields is bound to that of the symmetry algebra, such as gauge theories or non-linear sigma models, and the mass of the field is consequently forbidden by symmetry. Here, any massless field of symmetry group $G$ descended from a lattice of $N$ such copies must couple universally to all sites, barring a $G$ which satisfies the conditions for a third option for the $\varphi_i$. No meaningful clockwork mechanism exists. This is apparent, for instance, in the quiver of $SO(n)/SO(n-1)$ non-linear sigma models considered in \cite{Ahmed:2016viu}. In \cite{Ahmed:2016viu}, it is observed that higher order interactions of the Goldstone fields do not respect the putative remaining global symmetry (except in the case where $q=1$, i.e., where the Goldstone fields at each site shift equally).\footnote{Note that the interaction term (3.5) proposed in \cite{Ahmed:2016viu} to actually preserve the alleged symmetry factorizes by Baker-Campbell-Hausdorff into the original problematic interaction term in (3.2).}

The absence of a viable clockwork mechanism is manifest in the failure of explicit attempts to construct such a model for general $G$; we give one such example here. One might consider using a link field between adjacent sites that transforms under the direct product irrep $(r_1,r_2)$ of the direct product $G \times G$ of the left- and right-hand sites' symmetry groups. Let $r_1$ and $r_2$ be inequivalent unitary irreps of $G$.\footnote{This would be the generalization of using a link field with respective charges $1$ and $-q$ in the abelian case of section \ref{sec:discreteVector}.} Without loss of generality we may write the vev of the link field as a matrix $V$ transforming as
\begin{equation}
V \to \rho_{r_1}(g_1) V \rho_{r_2}^*(g_2^{-1}), (g_1,g_2) \in G \times G
\end{equation}
where $\rho_{r_1}(g)$ is the matrix representation $r_1$ of the given element $g$, and similarly for $r_2$. If the vev is invariant under the diagonal subgroup of $G \times G$,
\begin{equation}
V = \rho_{r_1}(g) V \rho_{r_2}^*(g^{-1}), \forall g \in G.
\end{equation}
Schur's lemma states that the only solution to the above equation, if $r_1$ and $r_2$ are inequivalent, is $V=0$. A non-zero vev of such a link field does not preserve the diagonal symmetry group.

However, in the case of $G=U(1)$, we may break to subgroups other than the diagonal. This freedom is exploited by the original clockwork models \cite{Choi:2015fiu,Kaplan:2015fuy}. The cyclic groups $\zz_q, q \in \nn$ (comprising respectively the $q$th roots of unity if $U(1)$ is multiplication on the set of unit modulus complex numbers) are normal subgroups $\zz_q \triangleleft U(1)$, and moreover the quotient group $U(1)/\zz_q$ is isomorphic to $U(1)$. So, in addition to the isomorphism or the map to the identity, we may choose, as a third option, the homomorphism
\begin{equation}
\varphi_i : U(1) \to \frac{U(1)}{\zz_{q_i}} .
\end{equation}
Writing everything as unit modulus complex numbers, we identify the most general $U(1) < U(1)^N$ as
\begin{equation}
\lbrace (\varphi_1(g),\varphi_2(g),\ldots,\varphi_N(g)) | g \in G \rbrace = \lbrace (e^{i q_1 \alpha},e^{i q_2 \alpha},\ldots, e^{i q_N \alpha} ) | 0 \leq \alpha < 2 \pi, q_i \in \nn \rbrace .
\label{eq:asymgroup}
\end{equation}

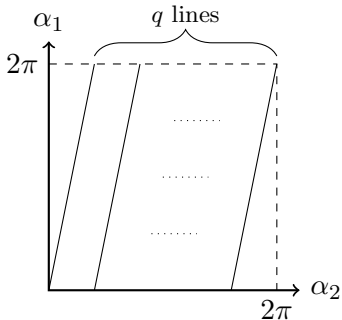
\begin{figure}
\centering
\begin{tikzpicture}[scale=1.5]
    \draw [<->,thick] (0,2.2) node (yaxis) [above] {$\alpha_1$}
        |- (2.2,0) node (xaxis) [right] {$\alpha_2$};
    \draw (0,0) -- (0.4,2);
    \draw (0.4,0) -- (0.8,2);
    \draw (1.6,0) -- (2,2);
    \draw [dotted] (1.0,1.0) -- (1.4,1.0);
    \draw [dotted] (0.9,0.5) -- (1.3,0.5);
    \draw [dotted] (1.1,1.5) -- (1.5,1.5);
    \coordinate (e) at (2,2);
    \draw[dashed] (yaxis |- e) node[left] {$2 \pi$}
        -| (xaxis -| e) node[below] {$2 \pi$};
    \draw [decorate,decoration={brace,amplitude=10pt},xshift=0pt,yshift=2pt] (0.4,2.0) -- (2.0,2.0) node [black,midway,yshift=15pt] {\footnotesize $q$ lines}; 
\end{tikzpicture}
\caption{The elements of the subgroup (\ref{eq:asymgroup}) highlighted in solid black on the toroidal $U(1)^2$ group manifold. Vertically level points on the top and bottom edges of the square are the same, as are the horizontally level points along the left and right edges of the square. See also the similar depictions in \cite{Choi:2015fiu}. \label{fig:windingtorus}}
\end{figure}

We highlight the specific $U(1)<U(1)^2$ subgroup of $q_1=q$, $q_2=1$, on the $U(1)^2$ group manifold in figure \ref{fig:windingtorus}. This symmetry breaking structure is realized concretely in a two-site clockwork model. The lagrangian,
\begin{equation}
\mathcal{L} = \abs{\partial \phi_1}^2 + \abs{\partial \phi_2}^2 + \epsilon \phi_1^\dagger \phi_2^q,
\end{equation}
has, when $\epsilon = 0$, a $U(1)\times U(1)$ symmetry with (w.l.o.g) unit charges such that $\phi_1 \to e^{i \alpha_1} \phi_1$ and $\phi_2 \to e^{i \alpha_2} \phi_2$ under the action of group element $(e^{i \alpha_1},e^{i \alpha_2})$. When $\epsilon \neq 0$, the potential term effects an explicit breaking to the subgroup where $\alpha_1 = q \alpha_2$, which is precisely the subgroup depicted in figure \ref{fig:windingtorus} and described in (\ref{eq:asymgroup}).

\subsection{The continuum limit}

Not all of the above symmetry breaking patterns can originate from the deconstruction of a higher dimensional theory. Smoothness considerations will in fact allow us to constrain the behavior of the resulting $\varphi_i$ even further for Lie group $G$.

To understand the `large $N$' or continuum limit of $G < G^N$, consider a trivial principal fiber bundle $G \times M$, where we have a copy of the Lie group --- a fiber $G_x$ --- for each point on the manifold $x \in M$. To find a subgroup $G < G \times M$, we effectively wish to find a set of sections $\varphi(g,x)$ of $G \times M$, which, for each $g \in G$, identify a corresponding element of $G_x$, for all $x \in M$. As in the discrete case above, we require that they respect pointwise group composition $\varphi(g,x)\varphi(g^\prime,x)=\varphi(g g^\prime,x), \forall x \in M, \forall g,g^\prime \in G$, and thus $\varphi(\cdot,x)$ must be a homomorphism from $G$ to a subgroup $G^\prime_x < G_x$.

As before, barring a $G$ that satisfies the conditions for a third option (such as a $G$ that is a non-trivial semidirect product), $\varphi(\cdot,x)$ must be either an isomorphism or a map to the trivial group. Continuity of $\varphi$ alone requires it must be exclusively one or the other for all $x \in M$, and the requirement that each section be distinct means it is an isomorphism for all $x$. Similar continuity arguments require, for $G=U(1)$, that $\varphi(\cdot,x) : U(1) \to U(1)/\zz_q$ with fixed $q$ for all $x \in M$, and requiring that each section be distinct sets $q=1$.

In the particular case of warped extra dimensions, where $G$ is either a $U(1)$ or $SU(n)$ 4D gauge group and $M = S_1/\zz_2$, it is a consequence that the profile of the zero mode gauge boson is flat in the bulk. (As can be seen by restricting the maps $\varphi$ to the global part of the 4D gauge transformation at each point in $M$.) In other words, the subset of 5D gauge transformations which preserve the $y$-dependence of the zero mode $A_\mu(x,y)= \tilde A^0_\mu(x) \psi^0(y)$ can only form the corresponding 4D gauge group if $\psi(y)$ is a constant. It is worth considering what happens in attempts to localize the zero mode of gauge bosons. In the non-abelian Yang Mills theory \cite{Batell:2006dp}, the localized zero mode's couplings to other fields, such as other KK modes and brane localized matter, are not gauge invariant. In the case of a $U(1)$ 5D gauge group \cite{Batell:2005wa}, a localized zero mode nevertheless transforms under an $\rr$ 4D gauge group, and the resulting 4D theory is nevertheless well behaved. This effect is seen in the (well-behaved) theory of section \ref{sec:vectorFlat}, where charges of brane localized matter under the zero mode's gauge symmetry are generally non-integer (see (\ref{eq:4dVectorBraneCoupling})) --- \emph{i.e.} the matter comes in representations of $\rr$, not $U(1)$. However, as these are all projective representations of $U(1)$, we are free to treat the low energy theory (of the zero mode and its couplings to matter) as a 4D $U(1)$ gauge theory.

\section{Deconstructing gravitational extra dimensions}
\label{app:graviton}

In this section, we briefly illustrate, following \cite{ArkaniHamed:2003vb}, how the deconstruction of a gravitational extra dimension leads
to the discrete, unclockworked, scenario of section \ref{sec:discreteGraviton} -- regardless of the choice of metric.
We consider perturbations around the geometry defined by eq.(\ref{eq:metric}), of the form
\begin{equation}
	ds^2 = g_{MN} dx^M dx^N = X(y) {\tilde g}_{\mu \nu} dx^\mu dx^\nu + Y(y) dy^2 \ ,
\end{equation}
where ${\tilde g}_{\mu \nu} = \eta_{\mu \nu} + h_{\mu \nu}$.
The 5D Einstein-Hilbert action is then given by
\begin{equation}
\begin{split}
	S_{5, {\rm EH}} & = \frac{M_5^3}{2} \int d^4 x dy \sqrt{|g|} R_5 [g] \\
				& = \frac{M_5^3}{2} \int d^4 x dy \sqrt{|{\tilde g}|} \Biggl( X \sqrt{Y} R_4 [{\tilde g}] \\
				& + \frac{1}{4} X^{-2} Y^{-1} ({\tilde g}^{\mu \nu} {\tilde g}^{\alpha \beta} - {\tilde g}^{\mu \alpha} {\tilde g}^{\nu \beta})
					\partial_y (X {\tilde g}_{\mu \nu}) \partial_y (X {\tilde g}_{\alpha \beta}) \Biggr) \ .
\end{split}
\label{eq:5DEH}
\end{equation}

Expanding eq.(\ref{eq:5DEH}) to quadratic order in $h_{\mu \nu}$, and writing $h_{\mu \nu}$ as a sum over KK-modes as usual,
$h_{\mu \nu} = \sum_{n = 0}^\infty \varphi_n (y) h_{\mu \nu}^{(n)} (x)$, one can find the corresponding equations of motion and boundary conditions.
In particular, a massless mode is present in the spectrum, whose profile is a constant
(to preserve the normalization of the massless mode's self interactions, we take $\varphi_0 = 1$).
The first term in eq.(\ref{eq:5DEH}) then defines the effective 4D Planck scale, given by
\begin{equation}
	M_{Pl}^2 = M_5^3 \int_{y=0}^{y=\pi R} dy X(y) \sqrt{Y(y)} \ .
\end{equation}

When deconstructed, the first term in the second equality of eq.(\ref{eq:5DEH}) leads to a term of the form
\begin{equation}
	{\mathcal L}_{4} \supset \sum_{j=0}^N \sqrt{|{\tilde g}_j|} \frac{M_5^3}{2} a X_j \sqrt{Y_j} R_4 [{\tilde g}_j] \\
				\equiv \sum_{j=0}^N \sqrt{|{\tilde g}_j|} \frac{M_j^2}{2} R_4 [{\tilde g}_j] \ ,
\label{eq:EHdeconstruction}
\end{equation}
where ${\tilde g}_{j \mu \nu} = \eta_{\mu \nu} + h_{j \mu \nu}$, and $M_j$ corresponds to the effective 4D Planck scale at site $j$, given by
\begin{equation}
	M_j^2 = M_5^3 a X_j \sqrt{Y_j} \ .
\label{eq:MPlsites}
\end{equation}
For instance, in a linear dilaton background of the form $X(y)=Y(y)=e^{-4ky}$, $M_j = (M_5^3 a)^{1/2} e^{-3 k j a}$,
whereas in a Randall-Sundrum geometry, $X(y)=e^{-2ky}$, $Y(y)$=1, one finds $M_j = (M_5^3 a)^{1/2} e^{-k j a}$, in agreement with~\cite{Randall:2005me}.
In both cases, the effective Planck scale on a given site depends exponentially on the position of the site.

Moreover, upon deconstruction, the second term in the second equality of eq.(\ref{eq:5DEH}) leads to a mass term of the form (expanding up to ${\mathcal O} (h^2)$)
\begin{equation}
	{\mathcal L}_{4} \supset \frac{1}{2} \sum_{j=0}^{N-1} \frac{M_j^2}{4 a^2} \frac{X_j}{Y_j} \left\{
					(h_{j} - h_{j+1})^2 - (h_{j \mu \nu} - h_{j+1 \mu \nu})^2 \right\} \ ,
\end{equation}
which corresponds precisely to eq.(\ref{eq:massTermGravLin}), with mass-squared parameters
\begin{equation}
	m_j^2 = \frac{1}{a^2} \frac{X_j}{Y_j} \ .
\end{equation}
The deconstructed theory is therefore identical to the discrete 4D scenario described in section~\ref{sec:discreteGraviton},
in which no clockwork graviton arises.

\end{document}